\documentclass[12pt, draftclsnofoot, onecolumn]{IEEEtran}
\pdfoutput=1
\usepackage{amssymb}
\usepackage{amsmath}
\usepackage{algorithmic}
\usepackage{algorithm}
\usepackage{cite}
\usepackage{url}
\usepackage{xcolor}
\usepackage{color} 
\usepackage{cite,graphicx,amsmath,amssymb}
\usepackage{subfigure}
\usepackage{citesort}
\usepackage{fancyhdr}
\usepackage{mdwmath}
\usepackage{mdwtab}
\usepackage{caption}
\usepackage{amsthm}
\usepackage{setspace}
\usepackage{xspace}
\usepackage{graphicx}
\usepackage[acronym,nogroupskip,nonumberlist,nopostdot]{glossaries}
\makeglossaries
\newacronym{aod}{AOD}{angle of departure}
\newacronym{aoa}{AOA}{angle of arrival}
\newacronym{snr}{SNR}{signal-to-noise ratio}
\newacronym{6g}{6G}{sixth-generation}
\newacronym{5g}{5G}{fifth-generation}
\newacronym{sre}{SRE}{smart radio environment}
\newacronym{stars}{STAR-RISs}{simultaneously transmitting and reflecting reconfigurable intelligent surfaces}
\newacronym{star}{STAR-RIS}{simultaneously transmitting and reflecting reconfigurable intelligent surface}
\newacronym{mimo}{MIMO}{multiple-input multiple-multiple}
\newacronym{cscg}{CSCG}{circularly symmetric complex Gaussian}
\newacronym{los}{LoS}{line-of-sight}
\newacronym{em}{EM}{electromagnetic}

\DeclareMathOperator{\diag}{diag}
\DeclareMathOperator{\Diag}{Diag}
\DeclareMathOperator{\rank}{Rank}
\DeclareMathOperator{\tr}{Tr}

\newcommand{\ie}{\text{i}.\text{e}.}
\newcommand{\eg}{\text{e}.\text{g}.}

\newcommand{\bm}[1]{\mbox{\boldmath{$#1$}}}

\newcommand{\sixg}{\gls{6g}\xspace}
\newcommand{\fiveg}{\gls{5g}\xspace}
\newcommand{\sre}{\gls{sre}\xspace}

\newcommand{\starriss}{\gls{stars}\xspace}
\newcommand{\mimo}{\gls{mimo}\xspace}
\newcommand{\cscg}{\gls{cscg}\xspace}
\newcommand{\los}{\gls{los}\xspace}
\newcommand{\emm}{\gls{em}\xspace}

\newtheorem{theorem}{Theorem}

\newtheorem{lemma}{Lemma}

\newtheorem{corollary}{Corollary}

\newtheorem{proposition}{Proposition}

\allowdisplaybreaks
\makeatletter
\def\ScaleIfNeeded{%
\ifdim\Gin@nat@width>\linewidth \linewidth \else \Gin@nat@width
\fi } \makeatother

\begin{document}

\title{Near-Field Beamforming for STAR-RIS Networks}

\author{

\vspace{-0.3cm}
Haochen~Li,
Yuanwei~Liu,~\IEEEmembership{Senior Member,~IEEE,}
Xidong~Mu,~\IEEEmembership{Member,~IEEE,}
Yue~Chen,~\IEEEmembership{Senior Member,~IEEE,}
Pan~Zhiwen,~\IEEEmembership{Member,~IEEE,}
Yonina~C.~Eldar,~\IEEEmembership{Fellow,~IEEE}

\thanks{This article has been submitted to the IEEE Global Communications Conference, in 2023~\cite{Li}.}
\thanks{H.~Li and Z.~Pan are with National Mobile Communications Research Laboratory, Southeast University, Nanjing 210096, China, and also with Purple Mountain Laboratories, Nanjing 211100, China (email:\{lihaochen, pzw\}@seu.edu.cn).}

\thanks{Y.~Liu, X.~Mu, and Y.~Chen are with Queen Mary University of London, London, UK (email:\{yuanwei.liu, xidong.mu, yue.chen\}@qmul.ac.uk).}

\thanks{Y.~C.~Eldar is with the Faculty of Math and CS, Weizmann Institute of Science, Rehovot 7632706, Israel (email:yonina.eldar@weizmann.ac.il)}

}

\maketitle
\vspace{-1.8cm}
\begin{abstract}
\vspace{-0.2cm}
Recently, simultaneously transmitting and reflecting reconfigurable intelligent surfaces (STAR-RISs) have received significant research interest. The employment of large STAR-RIS and high-frequency signaling inevitably make the near-field propagation dominant in wireless communications. In this work, a STAR-RIS aided near-field multiple-input multiple-multiple (MIMO) communication framework is proposed. A weighted sum rate maximization problem for the joint optimization of the active beamforming at the base station (BS) and the transmission/reflection-coefficients (TRCs) at the STAR-RIS is formulated. The non-convex problem is solved by a block coordinate descent (BCD)-based algorithm. In particular, under given STAR-RIS TRCs, the optimal active beamforming matrices are obtained by solving a convex quadratically constrained quadratic program. For given active beamforming matrices, two algorithms are suggested for optimizing the STAR-RIS TRCs: a penalty-based iterative (PEN) algorithm and an element-wise iterative (ELE) algorithm. The latter algorithm is conceived for STAR-RISs with a large number of elements. Numerical results illustrate that: i) near-field beamforming for STAR-RIS aided MIMO communications significantly improves the achieved weighted sum rate compared with far-field beamforming; ii) the near-field channels facilitated by the STAR-RIS provide enhanced \emph{degrees-of-freedom} and \emph{accessibility} for the multi-user MIMO system; and iii) the BCD-PEN algorithm achieves better performance than the BCD-ELE algorithm, while the latter has a significantly lower computational complexity. 
\end{abstract}\vspace{-0.4cm}

\section{Introduction}
\IEEEPARstart{T}o support emerging technologies such as virtual and augmented reality, autonomous vehicles, and digital replicas, the capacity of \sixg communication systems is required to be much superior to that of \fiveg communication systems~\cite{9349624}. To achieve this goal, one solution is to rely on new physical layer technologies. Recently, \starriss have been proposed~\cite{9690478,xu2022star,zuo2022joint}. By employing a large number of low-cost STAR elements, the wireless signal impinging on STAR-RISs can be split into transmitted and reflected sides, thus realizing a full-space \sre~\cite{9140329,9424177,liu2023simultaneously}.\\
\indent Given the above advantages, extensive research efforts have been devoted to utilizing STAR-RISs to achieve various objectives in wireless networks. These include reducing transmit power~\cite{9570143,liu2022simultaneously}, enhancing spectrum- and energy-efﬁciency~\cite{wang2022simultaneously,10052764}, and enlarging coverage area~\cite{wu2021coverage}. More speciﬁcally, the authors of~\cite{9570143} investigated joint beamforming design to minimize the transmit power of the base station (BS) under three proposed  STAR-RIS operating protocols. In~\cite{liu2022simultaneously}, the authors further studied the transmit power minimization problem with the practical coupled-phase constraint on \starriss. The work~\cite{wang2022simultaneously} proposed a power consumption model for STAR-RIS and maximized the spectrum- and energy-efficiency for wide-band \mimo communication systems in the THz band. The work~\cite{10052764} investigates the energy efficiency maximization for a multiple-input-single-output (MISO) STAR-RIS assisted non-orthogonal multiple access (NOMA) downlink network. In~\cite{wu2021coverage}, it was verified that STAR-RISs can extend the coverage of STAR-RIS aided NOMA and orthogonal multiple access (OMA) communication systems. The works above are limited to far-field communications where the conventional planar wave assumption holds and the \emm wave propagation is approximately modeled with far-field channels.\\ 
\indent With increased antenna/element number and high operating frequencies, wireless communications are likely to take place in the near-field region rather than the far-field region. The boundary between the two is the Rayleigh distance $\frac{2D^2}{\lambda}$, where $D$ is the array diameter and $\lambda$ is the wavelength of the carrier signal~\cite{7942128}. In the near field, the conventional planar wave assumption is invalid and the \emm wave propagation has to be accurately modeled with near-field channels~\cite{bjornson2021primer}. Compared to conventional planar-wave-based far-field wireless channels, the spherical-wave-based near-field wireless channels have several promising characteristics: i) {near-field beamfocusing:} owing to the different EM wave propagation model, the near-field channel contains both angle and distance information of users~\cite{9536436}. This enables the communication system to concentrate transmitted signals on a specific location and thus provides \emph{accessibility improvement} for multi-user communications; and ii) {high-rank \los \mimo channels:} thanks to the spherical propagation in the near field, the near-field \los channel is rank-sufficient~\cite{miller2019waves}. Compared with the low-rank far-field LoS channel, the near-field LoS channel can support more information streams for the multi-antenna users and thus provides \emph{degrees-of-freedom (DoFs) enhancement} for MIMO communications.\\
\indent A variety of studies have been recently devoted to exploring near-field characteristics in wireless communication systems. In~\cite{zhang2022near}, the authors proposed a distance-parameterized angular-domain sparse model for the near-field channels and developed a joint dictionary learning and sparse recovery algorithm to effectively estimate both the LoS and multi-path near-field channels. In~\cite{zhang2022beam}, near-ﬁeld beamfocusing for \mimo communications was investigated under three different types of MIMO antenna architectures. It was shown that, compared with conventional beamsteering, the utilization of near-field beamfocusing in multi-user MIMO communication systems leads to reduction of co-channel interference. The authors of~\cite{wu2022multiple} proposed using location information contained by the near-field channel to achieve multiplexing in the location domain. Physical layer security (PLS) of near-field communication systems was investigated in~\cite{zhang2023physical}. The work~\cite{zhang2022near1} investigated wireless power transfer (WPT) for charging mobile devices in the near-field of a dynamic metasurface antenna (DMA). With optimized DMA weights and beamformers, focused energy beams can be generated to improve the system energy transfer efficiency.\\
\indent Though there are a abundant of works on STAR-RIS aided communications and many studies concerning near-field communications, STAR-RIS aided communications in the near-field has not been investigated to date. STAR-RISs are extremely beneficial to be deployed for assisting millimeter-wave/Terahertz (mmWave/THz) communications via creating virtual \los links to overcome blockage~\cite{6995424}. Large size STAR-RISs comprising of many elements have to be employed to confront high pathloss attenuation at high frequency band~\cite{8901159}. As a result, the utilization of large STAR-RISs and the adoption of high-frequency signaling inherently result in the predominance of near-field propagation. For example, consider a STAR-RIS with an array diameter of $0.5$ meters operating at a frequency of $28$ GHz. In this case, the Rayleigh distance of the STAR-RIS array is approximately $47$ meters. Consequently, it is highly probable that users will be situated within the near-field region of the STAR-RIS. The fact that the favorable characteristics of near-field channels can be leveraged to improve the capacity of communication systems motivates the study of STAR-RIS aided near-field communications and the corresponding near-field beamforming design.\\
\indent In this work, we propose a STAR-RIS aided near-field MIMO communication framework for the first time, where a BS serves multiple users within the near-field region of the STAR-RIS. We formulate a weighted sum rate maximization problem to jointly optimize the active beamforming at the BS and the transmission/reflection-coefficients (TRCs) at the STAR-RIS, subject to a power constraint at the BS and a TRCs constraint at the STAR-RIS. The weighted sum rate maximization problem is solved by using a block coordinate descent (BCD)-based algorithm, where the active beamforming matrices at the BS and the TRCs at the STAR-RIS are alternatingly optimized. Specifically, under given fixed STAR-RIS TRCs, the optimal active beamforming matrices are obtained by solving a convex quadratically constrained quadratic program problem. Two algorithms are suggested for optimizing the STAR-RIS TRCs under given fixed active beamforming matrices. First, we propose the penalty-based iterative (PEN) algorithm to optimize STAR-RIS TRCs by invoking the successive convex approximation (SCA) technique. Then, for STAR-RIS with massive elements, we develop the low-complexity element-wise iterative (ELE) algorithm to optimize STAR-RIS TRCs by invoking the bisection method and  exhaustive search method. Our numerical results illustrate that: i) utilization of near-field beamforming in STAR-RIS aided MIMO communications leads to a substantial improvement in the achieved weighted sum rate with respect to far-field beamforming; ii) the near-field channels facilitated by the STAR-RIS provide enhanced \emph{degrees-of-freedom} and \emph{accessibility} for the multi-user MIMO system; and iii) the BCD-PEN algorithm outperforms the BCD-ELE algorithm but the BCD-ELE algorithm exhibits significantly lower computational complexity.\\
\indent The remainder of this paper is organized as follows: Section II introduces the system and channel models. In Section III, the optimization problem is reformulated and then solved by the developed BCD-PEN algorithm. In Section IV, a low-complexity BCD-ELE algorithm is proposed to solve the optimization problem when the number of STAR elements is large. Section V and Section VI present simulation results and conclusions, respectively.\\
\indent \textit{Notations:} Lowercase letters, lowercase bold letters, and capital bold letters denote scalars, vectors, and matrices, respectively. The $M \times K$ dimensional complex matrix space is denoted by $\mathbb{C}^{M \times K}$. The superscripts $(\cdot)^\mathrm{T}$, $(\cdot)^\mathrm{*}$, and  $(\cdot)^\mathrm{H}$ represent the operations of transpose, conjugate, and conjugate transpose, respectively. Symbols $\Diag\left( \mathbf{A} \right)$ and $\diag\left( \mathbf{a} \right)$ represent a vector whose elements are extracted from the main diagonal elements of the matrix $\mathbf{A}$ and a diagonal matrix with $\mathbf{a}$ on its main diagonal, respectively. Symbol $[\mathbf{A}]_{mn}$ is the element on the $m$-th row and $n$-th column of matrix $\mathbf{A}$. The distribution of a \cscg random vector with mean vector $\mathbf{a}$ and covariance matrix $\mathbf{A}$ is denoted as $\mathcal{C N}\left(\mathbf{a}, \mathbf{A}\right)$. Operation $\lfloor a\rfloor$ is the largest integer that is not greater than $a$, $\text{mod}(a,b)$ is the remainder of the Euclidean division of $a$ by $b$, and arg$(\cdot)$ means the extraction of phase information. Matrix $\mathbf{I}$ is an identity matrix with appropriate dimensions. Calligraphic letters represent sets, \eg, $\mathcal{A}$.\vspace{-0.2cm}
\section{System Model}
In this section, we present the system model for the investigated STAR-RIS aided near-field multi-user MIMO communication system and formulate the weighted sum rate maximization problem.\vspace{-0.4cm}
\subsection{System Model}
\begin{figure}[h]
\centering
\includegraphics[width=4in]{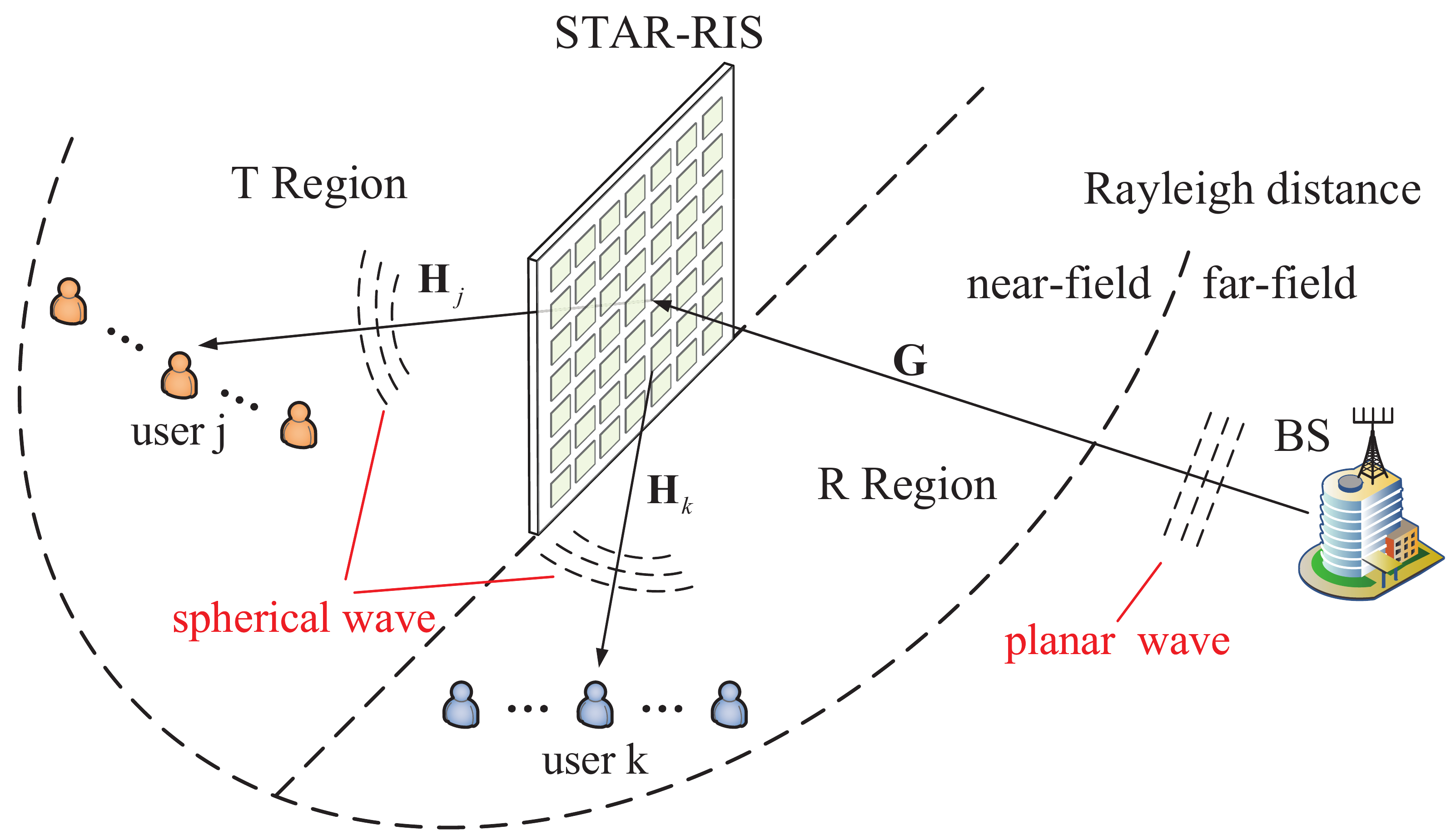}
\caption{STAR-RIS aided near-field MIMO communications.}
\label{system_model}
\end{figure}

We consider a STAR-RIS aided downlink multi-user MIMO system, where the BS equipped with an $M_b$-antenna uniform linear array (ULA) serves $K$ users equipped with $M$-antenna ULAs. As depicted in Fig.~\ref{system_model}, the BS-user links are blocked by obstacles, which is practical for systems operating under mmWave/THz bands since high-frequency signals are prone to absorption and susceptible to blockage. A STAR-RIS is deployed on the user side to create \los transmission and reflection links for the blocked users. It is assumed that the uniform planar array (UPA)-type STAR-RIS contains $N=N_y \times N_z$ elements. The users located in the transmission side and reflection side of the STAR-RIS are referred to as T users and R users, respectively. All users are collected in set $\mathcal{K}\buildrel \Delta \over = \left\{1,2, \cdots, K\right\}$. T users and R users are collected in subsets $\mathcal{K}_t\buildrel \Delta \over = \left\{1,2, \cdots, K_0\right\}$ and $\mathcal{K}_r\buildrel \Delta \over = \mathcal{K}\setminus\mathcal{K}_t$. \\
\indent Let $\mathbf{\Phi}_{i}=\diag\{\sqrt{\rho_{1}^i}e^{j\theta_{1}^i}, \sqrt{\rho_{2}^i}e^{j\theta_{2}^i}, \cdots, \sqrt{\rho_{N}^i}e^{j\theta_{N}^i}\}, \forall i\in\{t,r\}$ denote the TRC matrix of the STAR-RIS, where $\rho_n^t, \rho_n^r\in[0,1]$ are the amplitude coefficients for transmission and reflection and $\theta_n^t, \theta_n^r\in[0,2\pi)$ are the corresponding phase shifts introduced by the $n$-th elements. In this paper, we consider both energy splitting (ES) and mode switching (MS) protocols for the STAR-RIS and adopt the corresponding TRCs constraints proposed in~\cite{9570143}. For the ES protocol, all STAR elements operate in simultaneous transmission and reflection mode (T\&R mode). The set of constraints to the TRCs for the $n$-th STAR element is given by
\begin{equation}
\begin{aligned}
\mathcal{C}_{\text{ES}}=\left\{ {\begin{array}{*{20}{c}}
{\rho _n^t,\rho _n^r}&\vline& {\rho _n^t,\rho _n^r \in [0,1];\rho _n^t + \rho _n^r = 1}\\
{\theta _n^t,\theta _n^r}&\vline& {\theta _n^t,\theta _n^r \in [0,2\pi ), \forall n\in\mathcal{N}}
\end{array}} \right\},
\end{aligned}
\end{equation}
where $\mathcal{N} \buildrel \Delta \over=\{1,2,\cdots,N\}$. For the MS protocol, all STAR elements operate either in full transmission mode (T mode) or full reflection mode (R mode), and the corresponding set of constraints is
\begin{equation}
\begin{aligned}
\mathcal{C}_{\text{MS}}=\left\{ {\begin{array}{*{20}{c}}
{\rho _n^t,\rho _n^r}&\vline& {\rho _n^t,\rho _n^r \in \{0,1\};\rho _n^t + \rho _n^r = 1}\\
{\theta _n^t,\theta _n^r}&\vline& {\theta _n^t,\theta _n^r \in [0,2\pi ), \forall n\in\mathcal{N}}
\end{array}} \right\}.
\end{aligned}
\end{equation}\vspace{-0.4cm}
\vspace{-0.6cm}
\subsection{Channel Model}
As the STAR-RIS is deployed on the user side, the far-field channel model can be used for the BS-STAR-RIS link due to the relatively far distance. However, for the STAR-RIS-user link we use the near-field channel model because the users are located in the vicinity of the STAR-RIS. In the following, we introduce the far-field channel from the BS to the STAR-RIS and the near-field channel from the STAR-RIS to user $k$, which are denoted by $\mathbf{G}\in\mathbb{C}^{N \times M_b}$ and $\mathbf{H}_k\in\mathbb{C}^{M \times N}$, respectively.
\subsubsection{The far-field channel}
For the far-field BS-STAR-RIS channel, we adopt the widely used geometric channel model with $L$ scatterers
\begin{equation}
\begin{aligned}\label{H}
\mathbf{G}=\sqrt{\frac{\beta M_b N}{L}} \sum\nolimits_{l=1}^{L} \mathbf{e}_\text{STAR}\left(\varphi_{l}, \vartheta_{l}\right)\mathbf{e}^{\mathrm{H}}_\text{BS}\left(\gamma_{l}\right),
\end{aligned}
\end{equation}
where $L$ and $\beta$ denote the number of dominant paths and the path-loss coefficient, respectively. The array response vector for the UPA-type STAR-RIS is given by
\begin{equation}
\begin{aligned}
\mathbf{e}&_\text{STAR}\left(\varphi_l, \vartheta_l\right)=\\
&\left[1,e^{jk_cd_R\sin\varphi_l\sin\vartheta_l}, \cdots, e^{jk_c\left(N_x-1\right)d_R\sin\varphi_l\sin\vartheta_l}\right]^\mathrm{T}\otimes\left[1,e^{jk_cd_R\cos\vartheta_l}, \cdots, e^{jk_c\left(N_y-1\right)d_R\cos\vartheta_l}\right]^\mathrm{T},
\end{aligned}
\end{equation}
where $k_c=2\pi/\lambda_c$ and $d_R=\lambda_c$ denote the wave number and the STAR element spacing, respectively, $\lambda_c$ is the wavelength, $\varphi_l$ and $\vartheta_l$ represent the azimuth angle of arrival (AOA) and the elevation AOA associated with the STAR-RIS, respectively. The array response vector for the ULA at the BS is given by
\begin{equation}
\label{ULA}
\begin{aligned}
\mathbf{e}_\text{BS}\left(\gamma_l\right)&=\left[1,e^{jk_cd_B\cos\gamma_l},\cdots,e^{jk_c\left(M_b-1\right)d_B\cos\gamma_l}\right]^\mathrm{T},
\end{aligned}
\end{equation}
where $d_B=\lambda_c/2$ is the BS antenna spacing, and $\gamma_l$ represents the angle
of departure (AOD) associated with the BS.\vspace{-0.1cm}
\subsubsection{Near-field STAR-RIS-user \los channel}
\begin{figure}[h]
\centering
\includegraphics[width=2.5in]{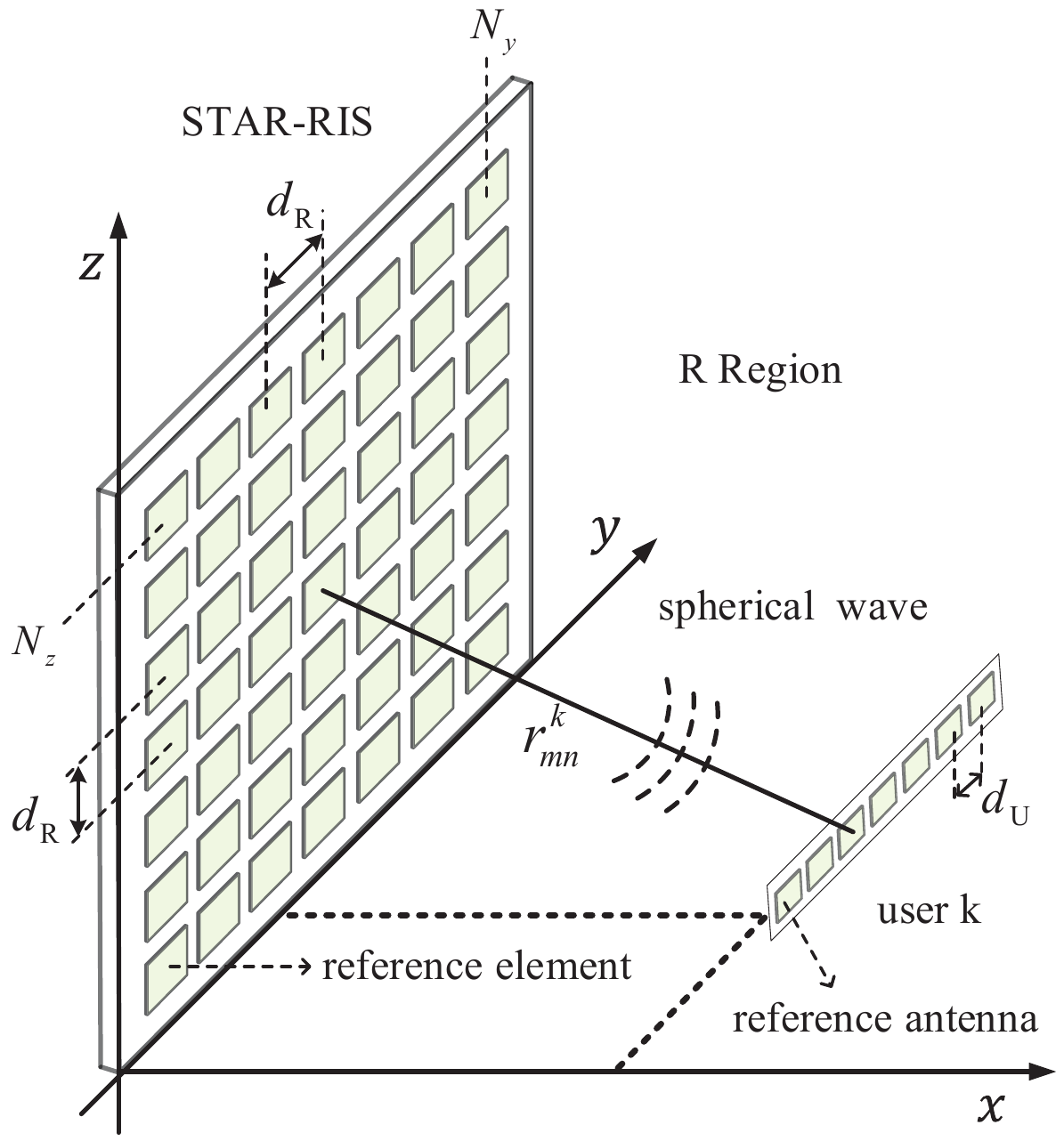}
\caption{The near-field \los channel.}
\label{fig_2}
\end{figure}
\vspace{-0.2cm}
For the near-field STAR-RIS-user \los channel, we consider a three-dimensional (3D) topology, where the Cartesian coordinate of the reference element of the STAR-RIS and the reference antenna of user $k$ are denoted by $\left(0, y_f, z_f\right)$ and $\left(x_k, y_k, 0\right)$, respectively. For simplicity, we assume that user $k$ is located in the R region of the STAR-RIS.\\
\indent Fig.~\ref{fig_2} shows the assumed setup of the STAR-RIS. The STAR-RIS is in the $YZ$-plane and all the users are in the $XY$-plane. If the STAR elements are indexed row by row from the bottom to the top, the Cartesian coordinate of the $n$-th STAR element, $n\in\mathcal{N}$, is given by 
\begin{equation}
\mathbf{p}_{n}=\left[0, i_{\mathrm{y}}\left(n\right)d_R+y_f, i_{\mathrm{z}}\left(n\right)d_R+z_f\right]^{\mathrm{T}},
\end{equation}
where $i_{\mathrm{y}}\left(n\right) = \mathrm{mod}\left(n-1,N_{\mathrm{y}}\right)$ and $i_{\mathrm{z}}\left(n\right) = \lfloor \left(n-1\right)/N_{\mathrm{y}}\rfloor$ are the column index and row index of the $n$-th element, respectively.\\
\indent Furthermore, we assume that the ULAs of all users are parallel to the $y$-axis. Thus, the Cartesian coordinate of the $m$-th antenna of user $k$, $m\in\mathcal{M}=\{1,2,\cdots,M\}$, is given by 
\begin{equation}
\mathbf{u}_{m}^k=\left[x_k, \left(m-1\right)d_U+y_z, 0\right]^{\mathrm{T}},
\end{equation}
where $d_U=\lambda_c/2$ is the user antenna spacing.\\
\indent In the near-field, we assume the EM signals propagate in spherical wavefronts and undergo free space path-loss. Then, the near-field \los channel between the $n$-th STAR element and the $m$-th antenna of user $k$ is given by~\cite{zhang2022beam,cui2021near}
\begin{equation}\label{nearH}
\begin{aligned}
\left[\mathbf{H}_k\right]_{mn}=\alpha_{mn}^k\exp\left(-j2\pi r_{mn}^k/\lambda_c\right),
\end{aligned}
\end{equation}
where $\alpha_{mn}^k=\frac{\lambda_c}{4\pi r_{mn}^k}$ and $r_{mn}^k=\left\|\mathbf{u}_{m}^k-\mathbf{p}_{n}\right\|_2$ represent the free space path-loss coefficient and the distance between the $m$-th antenna of user $k$ and the $n$-th STAR element, respectively.\\
\indent {The near-field STAR-RIS-user \los channel contains both the angle and distance information of users since the term $r_{mn}^k$ can be rewritten as follows:
\begin{equation}
\begin{aligned}
r_{mn}^k\left(\hat{\theta}_k,\hat{r}_k\right)=\left\|\left[\hat{r}_k\cos\hat{\theta}_k, \left(m-1\right)d_U+\hat{r}_k\sin\hat{\theta}_k, 0\right]^{\mathrm{T}}-\mathbf{p}_{n}\right\|_2,
\end{aligned}
\end{equation}
where $\hat{\theta}_k=\arctan\left(y_k/x_k\right)$ and $\hat{r}_k=r_{11}^k$ are the angle and distance of user $k$, respectively. This allows the STAR-RIS to transmit/reflect the intended signal towards not only a specific angle but also a specific distance, bringing about} accessibility enhancement {for multi-user communication systems.}\\
\indent When the far-ﬁeld approximation holds, the EM wave is approximately a parallel wave. In this case, as shown in~\cite{zhang2022near}, the phase shifts for elements of the STAR-RIS-user channel are linear along the user array, the rows and columns of the STAR-RIS, respectively. The path-loss for elements of the STAR-RIS-user channel is the same. Then, the near-field \los channel between the STAR-RIS and the $k$-th user degenerates into the far-field \los channel which can be expressed as \vspace{-0.2cm}
\begin{equation}\label{farH}
\begin{aligned}
\mathbf{H}_k^{\text{far}}=\sqrt{\beta_k M N} \mathbf{e}_\text{user}\left(\gamma_{k}\right) \mathbf{e}^{\mathrm{H}}_\text{STAR}\left(\varphi_{k}, \vartheta_{k}\right) ,
\end{aligned}
\vspace{-0.2cm}
\end{equation}

\noindent where $\gamma_{k}$ is the AoA associated with the $k$-th user, $\varphi_{k}$ and $\vartheta_{k}$ represent the azimuth AoD and the elevation AoD associated with the STAR-RIS, respectively, $\gamma_{k}$, $\varphi_{k}$, and $\vartheta_{k}$ are decided by the location of user $k$ with respect to the STAR-RIS, and $\beta_k$ is the path-loss coefficient between the STAR-RIS and user $k$. The array response vector for the ULA of users at the azimuth angle of $\gamma_k$ is similar to that in~\eqref{ULA}.\\
\indent In contrast to conventional far-field STAR-RIS-user \los \mimo channel in~\eqref{farH} being of rank-one, the near-field STAR-RIS-user \los \mimo channel in~\eqref{nearH} can be represented by a higher matrix due to the spherical-wave-based propagation model. {Specifically, as shown in~\cite{miller2000communicating}, the rank of the near-field LoS channel can be approximated with the DoFs of the prolate spheroidal wave functions, which significantly improves with decreased distance.} Thus, the near-field STAR-RIS-user \los \mimo channel can bear multiple information streams and provide a noticeable spatial DoFs improvement for MIMO communication systems.
\subsection{Problem Formulation}

The DoFs of the prolate spheroidal wave functions, which greatly increases with increasing distance, can be used to estimate the DoFs of the near-field LoS channel.
The signal sent by the BS can be expressed as
\begin{equation}
\begin{aligned}
\mathbf{x}=\sum\nolimits_{k\in\mathcal{K}}\mathbf{W}_k\mathbf{s}_k,
\end{aligned}
\end{equation}
where $\mathbf{s}_k\in\mathbb{C}^{M \times 1}$ and $\mathbf{W}_k\in\mathbb{C}^{M_b \times M}$ represent the symbol vector and the beamformer for user $k$, respectively. We assume that $\mathbb{E}\left\{\mathbf{s}_k(\mathbf{s}_k)^\mathrm{H}\right\}=\mathbf{I}$ and $\mathbb{E}\left\{\mathbf{s}_k(\mathbf{s}_{l})^\mathrm{H}\right\}=\mathbf{0}$, for $k\ne l$. The received signal vector at user $k$, $\forall k \in \mathcal{K}_i$, $\forall i \in \{t,r\}$, is given by
\begin{equation}\label{y}
\begin{aligned}
\mathbf{y}_k&=\mathbf{H}_k\mathbf{\Phi}_i\mathbf{G}\mathbf{x}+\mathbf{n}_k=\mathbf{H}_k\mathbf{\Phi}_i\mathbf{G}\mathbf{W}_k\mathbf{s}_k+\sum\nolimits_{l\in\mathcal{K}\setminus k}\mathbf{H}_k\mathbf{\Phi}_i\mathbf{G}\mathbf{W}_{l}\mathbf{s}_{l}+\mathbf{n}_k,
\end{aligned}
\end{equation}
where $\mathbf{n}_k\in\mathbb{C}^{M \times 1}$ is a noise vector which we assume is normal with distribution $\mathcal{CN}\left(\mathbf{0}, \sigma^2\mathbf{I}\right)$. The achievable data rate of user $k$, $\forall k \in \mathcal{K}_i$, $\forall i \in \{t,r\}$, is given by
\begin{equation}\label{R}
\begin{aligned}
R_k\left(\mathbf{W},\mathbf{\Phi}\right)=\log\left|\mathbf{I}+\mathbf{H}_k\mathbf{\Phi}_i\mathbf{G}\mathbf{W}_k\mathbf{W}_k^\mathrm{H}\mathbf{G}^\mathrm{H}\mathbf{\Phi}_i^\mathrm{H}\mathbf{H}_k^\mathrm{H}\mathbf{J}_k^{-1}\right|,
\end{aligned}
\end{equation}
where $\mathbf{W}\buildrel \Delta \over = \{\mathbf{W}_k, \forall k\in\mathcal{K}\}$, $\mathbf{\Phi}\buildrel \Delta \over = \{\mathbf{\Phi}_t,\mathbf{\Phi}_r\}$, and $\mathbf{J}_k$ is the interference-plus-noise covariance matrix
\begin{equation}\label{J}
\begin{aligned}
\mathbf{J}_k=\sum\nolimits_{l\in\mathcal{K}\setminus k}\mathbf{H}_k\mathbf{\Phi}_i\mathbf{G}\mathbf{W}_{l}\mathbf{W}_{l}^\mathrm{H}\mathbf{G}^\mathrm{H}\mathbf{\Phi}_i^\mathrm{H}\mathbf{H}_k^\mathrm{H}+\sigma^2\mathbf{I}.
\end{aligned}
\end{equation}

In this work, we focus on maximizing the weighted sum rate of the users by optimizing the  transmit beamforming matrices $\mathbf{W}$ and the TRC matrices $\mathbf{\Phi}$ subject to the total transmit power limit of the BS and the transmission-reflection constraint of STAR-RIS. The corresponding optimization problem can be formulated as:
\begin{subequations}\label{P1}
\begin{align}
\label{OBP1}\mathop {\max }\limits_{ {{\mathbf{W}},{\mathbf{\Phi}}} } \;\;&\sum\nolimits_{k\in\mathcal{K}}\eta_k R_k\left(\mathbf{W},\mathbf{\Phi}\right)\\[-0.2cm]
\label{Phi}{\rm{s.t.}}\;\;& {\theta}_n^i, {\rho}_n^i\in \mathcal{C}_{\mathcal{X}},\forall i \in \left\{ {t,r} \right\}, n \in \mathcal{N}\\[-0.2cm]
\label{power}&\sum\nolimits_{k\in\mathcal{K}}\|\mathbf{W}_k\|_{\text{F}}^{2}\le P,
\end{align}
\end{subequations}
where the constraint \eqref{Phi} is the TRC constraints for the STAR elements, with $\mathcal{X}\in\{\text{ES}, \text{MS}\}$ indicating the employed STAR-RIS operating protocol. The constraint \eqref{power} is the power constraint of the BS, with $P$ denoting the maximum transmit power of the BS. Here $\eta_k$ denotes the access priority for user $k$. In the following, we solve the non-convex problem~\eqref{P1} using the BCD method.

We note that when all STAR elements work in R mode and all users are located in the R region, the optimization problem \eqref{P1} for STAR-RISs degenerates into the optimization problem for conventional RISs. Thanks to the generality of the studied problem, the proposed joint beamforming algorithm can also serve as a general solution to the weighted sum rate maximization problem for conventional RIS-aided systems in the near field scenario.
\section{Proposed BCD-PEN Based Solutions}
In this section, the BCD-PEN algorithm is proposed to solve~\eqref{P1} under both ES and MS protocols.\vspace{-0.4cm}
\subsection{Optimization Problem Reformulation}
{Because of the coupling impact between TRC matrices and transmit beamforming matrices,~\eqref{P1} is challenging to solve. In this section, we capitalise on the connection between data rate and mean-square error (MSE) of the optimal combining matrix to reformulate the optimization problem.} 
Assuming the receivers adopt linear combining matrices, the estimated signal vector of the user $k\in\mathcal{K}$ is given by\vspace{-0.1cm}
\begin{equation}
\begin{aligned}
\tilde{\mathbf{s}}_k=\mathbf{U}_k^\mathrm{H}\mathbf{y}_k,
\end{aligned}
\vspace{-0.1cm}
\end{equation}
where ${\mathbf{U}}_k\in\mathbb{C}^{M \times M}$ is the combining matrix for the user $k$.\\
\indent Then, the MSE matrix of the user $k\in\mathcal{K}$ is given by
\begin{equation}\label{EMSE}
\begin{aligned}
{\mathbf{E}}_k^{\text{MSE}}=&\mathbb{E}\{\left(\tilde{\mathbf{s}}_k-{\mathbf{s}}_k\right)\left(\tilde{\mathbf{s}}_k-{\mathbf{s}}_k\right)^\mathrm{H}\}\\
=&\left(\mathbf{U}_k^\mathrm{H}\bar{\mathbf{H}}_k\mathbf{W}_k-\mathbf{I}\right)\left(\mathbf{U}_k^\mathrm{H}\bar{\mathbf{H}}_k\mathbf{W}_k-\mathbf{I}\right)^\mathrm{H}+\mathbf{U}_k^\mathrm{H}\bar{\mathbf{H}}_k\mathbf{W}_{\bar{k}}\mathbf{W}_{\bar{k}}^\mathrm{H}\bar{\mathbf{H}}_k^\mathrm{H}\mathbf{U}_k+\sigma^2\mathbf{U}_k^\mathrm{H}\mathbf{U}_k,
\end{aligned}
\end{equation} 
where $\bar{\mathbf{H}}_k\in\mathbb{C}^{M \times M_b}={\mathbf{H}}_k{\mathbf{\Phi}}_i{\mathbf{G}}$ stands for the aggregation channel from the BS to the user $k, \forall k \in \mathcal{K}_i$, $\forall i \in \{t,r\}$.\\
\indent  We use the weighted minimum mean square error (WMMSE) method to convert \eqref{OBP1} into a more tractable form. \eqref{P1} can be reformulated as follows~\cite{shi2011iteratively}:
\begin{subequations}\label{P2}
\begin{align}
\label{OBP2}\mathop {\min }\limits_{ {{\mathbf{Z}},{\mathbf{U}},{\mathbf{W}},{\mathbf{\Phi}}} } \;\;&\sum\nolimits_{k\in\mathcal{K}}\eta_kf_k\left(\mathbf{Z},\mathbf{U},\mathbf{W},\mathbf{\Phi}\right)\\[-0.2cm]
{\rm{s.t.}}\;\;& \eqref{Phi},\eqref{power},
\end{align}
\end{subequations}
where $\mathbf{Z}=\{\mathbf{Z}_k\succeq \mathbf{0}, \forall k \in \mathcal{K}\}$ and $\mathbf{U}=\{\mathbf{U}_k, \forall k \in \mathcal{K}\}$ denote the set of auxiliary matrices and the set of combining matrices, respectively. The function $f_k\left(\mathbf{Z},\mathbf{U},\mathbf{W},\mathbf{\Phi}\right)$, $\forall k \in \mathcal{K}$, is given by
\begin{equation}\label{f}
\begin{aligned}
f_k\left(\mathbf{Z},\mathbf{U},\mathbf{W},\mathbf{\Phi}\right)=\log\left|\mathbf{Z}_k\right|-\tr\left(\mathbf{Z}_k\mathbf{E}_k^{\text{MSE}}\right)+M.
\end{aligned}
\end{equation}
\subsection{Proposed BCD algorithm for joint beamforming }
{The objective function of the reformulated optimization problem is more tractable since it is concave with respect to $\mathbf{U}$, $\mathbf{Z}$, and $\mathbf{W}$ for given TRC matrices $\mathbf{\Phi}$.} In the following, we solve~\eqref{P2} iteratively using the BCD method. To elaborate, the optimization variables are divided into four blocks, \ie, $\mathbf{U}$, $\mathbf{Z}$, $\mathbf{W}$, and $\mathbf{\Phi}$. In each iteration, we optimize the variables in one block while the other blocks remain constant.
\subsubsection{Subproblem with respect to $\mathbf{U}$}
Given $\mathbf{Z}$, $\mathbf{W}$ and $\mathbf{\Phi}$, the optimal $\mathbf{U}$ in \eqref{P2} can be obtained by solving ${{\partial {f_{k}}} \mathord{\left/
{\vphantom {{\partial {f_{k}}} {\partial {\mathbf{U}_k}}}} \right.\kern-\nulldelimiterspace} {\partial {\mathbf{U}_k}}}=\mathbf{0}$, for $\forall k \in \mathcal{K}$. The solution is
\begin{equation}\label{U}
\begin{aligned}
\mathbf{U}_k^{\text{opt}}=\left(\mathbf{J}_k+\bar{\mathbf{H}}_k\mathbf{W}_{{k}}\mathbf{W}_{{k}}^\mathrm{H}\bar{\mathbf{H}}_k^\mathrm{H}\right)^{-1}\bar{\mathbf{H}}_k\mathbf{W}_k.
\end{aligned}
\end{equation}
\subsubsection{Subproblem with respect to $\mathbf{Z}$}
Similarly, given $\mathbf{U}$, $\mathbf{W}$ and $\mathbf{\Phi}$, the optimal $\mathbf{Z}$ in~\eqref{P2} is obtained by solving ${{\partial {f_{k}}} \mathord{\left/
{\vphantom {{\partial {f_{k}}} {\partial {\mathbf{Z}_k}}}} \right.\kern-\nulldelimiterspace} {\partial {\mathbf{Z}_k}}}=\mathbf{0}$, for $ \forall k \in \mathcal{K}$, which leads to
\begin{equation}\label{Z}
\begin{aligned}
\mathbf{Z}_k^{\text{opt}}=\left({\mathbf{E}}_k^{\text{opt}}\right)^{-1},
\end{aligned}
\end{equation}
where ${\mathbf{E}}_k^{\text{opt}}$ is obtained by inserting $\mathbf{U}_k^{\text{opt}}$ into \eqref{EMSE}.\\
\indent Substituting \eqref{U} and \eqref{Z} into \eqref{f}, the objective function of~\eqref{P2} can be rewritten as
\begin{equation}
\begin{aligned}
f_k\left(\mathbf{Z}^{\text{opt}},\mathbf{U}^{\text{opt}},\mathbf{W},\mathbf{\Phi}\right)&=\log\left\|\left({\mathbf{E}}_k^{\text{opt}}\right)^{-1}\right\|\buildrel (a) \over=\log\left\|\mathbf{I}+\mathbf{W}_k^\mathrm{H}\bar{\mathbf{H}}_k^\mathrm{H}\mathbf{J}_k^{-1}\bar{\mathbf{H}}_k\mathbf{W}_k\right\|\\
&=\log\left\|\mathbf{I}+\bar{\mathbf{H}}_k\mathbf{W}_k\mathbf{W}_k^\mathrm{H}\bar{\mathbf{H}}_k^\mathrm{H}\mathbf{J}_k^{-1}\right\|=R_k\left(\mathbf{W},\mathbf{\Phi}\right),
\end{aligned}
\end{equation}
where equality (a) comes from the Woodbury matrix identity. This confirms that~\eqref{P1} and~\eqref{P2} are equivalent with respect to solutions of $\mathbf{W}$ and $\mathbf{\Phi}$.
\subsubsection{Subproblem with respect to $\mathbf{W}$}
Given $\mathbf{Z}$, $\mathbf{U}$ and $\mathbf{\Phi}$, the weighted MSE  minimization problem \eqref{P2} can be transformed into the active beamforming problem as
\begin{subequations}\label{P3}
\begin{align}
\mathop {\min }\limits_{ {{\mathbf{W}}} } \;\;&\sum\nolimits_{k\in\mathcal{K}}g\left(\mathbf{W}\right)\\
{\rm{s.t.}}\;\;& \eqref{power},
\end{align}
\end{subequations}
where the objective function of~\eqref{P3} is given by 
\begin{equation}
\begin{aligned}
g\left(\mathbf{W}\right)=\sum\nolimits_{k\in\mathcal{K}}\tr\left(\mathbf{W}_k^\mathrm{H}\mathbf{A}\mathbf{W}_k\right)-\sum\nolimits_{k\in\mathcal{K}}2\tr\left(\Re\left(\mathbf{B}_k\mathbf{W}_k\right)\right),
\end{aligned}
\end{equation}
wherein $\mathbf{A}$ and $\mathbf{B}_k$ are denoted by $\mathbf{A}=\sum\nolimits_{l\in\mathcal{K}}\eta_l\bar{\mathbf{H}}_l^\mathrm{H}\mathbf{U}_l\mathbf{Z}_l\mathbf{U}_l^\mathrm{H}\bar{\mathbf{H}}_l$ and $\mathbf{B}_k=\eta_k\mathbf{Z}_k\mathbf{U}_k^\mathrm{H}\bar{\mathbf{H}}_k$, respectively. \eqref{P3} is a quadratically constrained quadratic program problem and can be solved by CVX~\cite{grant2014cvx}. {Using the Lagrangian dual decomposition method~\cite{pan2020multicell}, the closed form solution of~\eqref{P3} can be obtained with reduced complexity.}
\subsubsection{Subproblem with respect to $\mathbf{\Phi}$}
Finally, for given $\mathbf{Z}$, $\mathbf{U}$ and $\mathbf{W}$, substitute $\mathbf{E}_k^{\text{MES}}$ into~\eqref{f}. Then the TRC matrices optimization problem is formulated as
\begin{subequations}\label{P4}
\begin{align}
\mathop {\min }\limits_{ {{\mathbf{\Phi}}} } \;\;&\sum\nolimits_{k\in\mathcal{K}}{g}_k\left(\mathbf{\Phi}\right)\\[-0.2cm]
{\rm{s.t.}}\;\;& \eqref{Phi},
\end{align}
\end{subequations}
where the objective function~\eqref{P4} is given by
\begin{equation}
\begin{aligned}
{g}_k\left(\mathbf{\Phi}\right)=\tr\left(\mathbf{\Phi}_i^\mathrm{H}\mathbf{C}_k\mathbf{\Phi}_i\mathbf{D}-2\Re\left(\mathbf{E}_k\mathbf{\Phi}_i\right)\right), \forall k \in \mathcal{K}_i, \forall i \in \{t,r\},
\end{aligned}
\end{equation}
wherein $\mathbf{C}_k$, $\mathbf{D}$, and $\mathbf{E}_k$ are denoted by $\mathbf{C}_k=\eta_k\mathbf{H}_k^\mathrm{H}\mathbf{U}_k\mathbf{Z}_k\mathbf{U}_k^\mathrm{H}\mathbf{H}_k$, $\mathbf{D}=\mathbf{G}\left(\sum_{l\in\mathcal{K}}\mathbf{W}_l\mathbf{W}_l^\mathrm{H}\right)\mathbf{G}^\mathrm{H}$, and $\mathbf{E}_k=\eta_k\mathbf{G}\mathbf{W}_k\mathbf{Z}_k\mathbf{U}_k^\mathrm{H}{\mathbf{H}}_k, \forall k \in \mathcal{K}$, respectively. \eqref{P4} is non-convex due to the non-convex TRC constraints. We propose the PEN algorithm to solve this problem for STAR-RISs under both ES and MS protocols.
\subsection{Penalty-Based TRC Matrices Design Algorithm for ES and MS}
Using the matrix identities, $\forall k \in \mathcal{K}_i, \forall i \in \{t,r\}$,
\begin{equation}
\begin{aligned}
\tr\left(\mathbf{\Phi}_i^\mathrm{H}\mathbf{C}_k\mathbf{\Phi}_i\mathbf{D}\right)=\mathbf{v}_i^\mathrm{H}\mathbf{F}_k\mathbf{v}_i, \quad \tr\left(\mathbf{\Phi}_i\mathbf{E}_k\right)=\mathbf{e}_k^\mathrm{T}\mathbf{v}_i,
\end{aligned}
\end{equation}
where $\mathbf{v}_{i}=\Diag\left(\mathbf{\Phi}_{i}\right)=\left[v^i_1, v^i_2, \cdots, v^i_N\right]^{\mathrm{T}}, \forall i \in \{t,r\}$  is the STAR-RIS TRC vector, $\mathbf{F}_k=\left(\mathbf{C}_k \odot \mathbf{D}^\mathrm{T}\right)$ and, $\mathbf{e}_{k}=\Diag\left(\mathbf{E}_{k}\right)$, the objective function of~\eqref{P4} can be reformulated as 
\begin{equation}\label{Gk}
\begin{aligned}
\sum\nolimits_{i\in \{t,r\}}\left(\mathbf{v}_i^\mathrm{H}\sum\nolimits_{k\in\mathcal{K}_i}\mathbf{F}_k\mathbf{v}_i-2\Re\left(\sum\nolimits_{k\in\mathcal{K}_i}\mathbf{e}_k^\mathrm{T}\mathbf{v}_i\right)\right)=\sum\nolimits_{i\in \{t,r\}}\left(\mathbf{v}_i^\mathrm{H}\mathbf{F}_i\mathbf{v}_i-2\Re\left(\mathbf{e}_i^\mathrm{T}\mathbf{v}_i\right)\right),
\end{aligned}
\end{equation}
where $\mathbf{v}\buildrel \Delta \over = \left\{\mathbf{v}_t, \mathbf{v}_r\right\}$, $\mathbf{F}_i=\sum_{k\in\mathcal{K}_i}\mathbf{F}_k$ and $\mathbf{e}_i=\sum_{k\in\mathcal{K}_i}\mathbf{e}_k$, $\forall i\in\{t,r\}$.

To facilitate the design, we define the augmented TRC vector $\bar{\mathbf{v}}_i=\left[\mathbf{v}_i^\mathrm{T} \ \sqrt{t_i}\right]^\mathrm{T}, \forall i\in\{t,r\}$, where $\{t_t, t_r\}$ $\in\mathbb{R}$ are auxiliary variables. Moreover, we define ${\mathbf{V}}_i=\bar{\mathbf{v}}_i\bar{\mathbf{v}}_i^\mathrm{H}, \forall i\in\{t,r\}$, which satisfies ${\mathbf{V}}_i\succeq 0$, $\rank\{{\mathbf{V}}\}_i=1$, and $\Diag\{{\mathbf{V}}_i\}={\bm{\rho}}_i$, where ${\bm{\rho}}_i=\left[\rho_{1}^i,\rho_{2}^i,\cdots,\rho_{N}^i,t^i\right]^\mathrm{T}$. We rewrite the quadratic term $\mathbf{v}_i^{\mathrm{H}}\mathbf{F}_i\mathbf{v}_i-2\Re\{\mathbf{e}_i^{\mathrm{T}}\mathbf{v}_i\}$ as follows:
\begin{equation}
\begin{aligned}
\mathbf{v}_i^{\mathrm{H}}\mathbf{F}_i\mathbf{v}_i-2\Re\{\mathbf{e}_i^{\mathrm{T}}\mathbf{v}_i\}=\tr\left({\mathbf{V}}_i\bar{\mathbf{F}}_i\right), \forall i \in \{t,r\},
\end{aligned}
\end{equation}
where 
\begin{equation}
\begin{aligned}
\bar{\mathbf{F}}_i=\begin{bmatrix}
\mathbf{F}_i & -\mathbf{e}_i^{*} \\
-\mathbf{e}_i^\mathrm{T} & 0 
\end{bmatrix}, \forall i \in \{t,r\}.
\end{aligned}
\end{equation}
Then,~\eqref{P4} can be equivalently written as
\begin{subequations}\label{P5}
\begin{align}
&\mathop {\min }\limits_{ {{\mathbf{V}},{\bm{\rho}}} } \;\;\sum\nolimits_{\forall i \in \{t,r\}}\tr\left({\mathbf{V}}_i\bar{\mathbf{F}}_i\right)  \\[-0.15cm]
\label{rank V}{\rm{s.t.}}\;\;&\rank\{{\mathbf{V}}_i\}=1, \forall i \in\{t,r\},\\[-0.15cm]
\label{amplitude1}&\Diag\{{\mathbf{V}}_i\}={\bm{\rho}}_i, \forall i \in\{t,r\},\\[-0.15cm]
\label{semi V}&{\mathbf{V}}_i\succeq 0, \forall i \in\{t,r\},\\[-0.15cm]
\label{amplitude2}&\rho_n^t, \rho_n^r \in [0,1], \rho_n^t+\rho_n^r=1, \forall n\in \mathcal{N},\\[-0.15cm]
\label{auxiliary}&t^i=1, \forall i \in\{t,r\},\\[-0.15cm]
\label{MS}&\rho_n^t, \rho_n^r \in\{0,1\}, \forall n \in \mathcal{N},
\end{align}
\end{subequations} 
where $\mathbf{V}\buildrel \Delta \over =\{{\mathbf{V}}_t,{\mathbf{V}}_r\}$ and $\bm{\rho}\buildrel \Delta \over =\{{\bm{\rho}}_t,{\bm{\rho}}_r\}$ are the sets of optimization variables for~\eqref{P5}. We note that constraints \eqref{rank V}-\eqref{auxiliary} are the TRC constraints for ES while constraint \eqref{MS} is only present for MS. Both the rank-one constraint \eqref{rank V} and the binary constraint \eqref{MS} are non-convex.
\subsubsection{Penalty-based algorithm for ES}
To tackle the rank-one constraint~\eqref{rank V}, we rewrite it as following:
\begin{equation}\label{nuclear}
\begin{aligned}
\left\|{\mathbf{V}}_i\right\|_*-\left\|{\mathbf{V}}_i\right\|_2=0, \forall i\in\{t,r\},
\end{aligned}
\end{equation}
where $\left\|{\mathbf{V}}_i\right\|_*$ and $\left\|{\mathbf{V}}_i\right\|_2$ denote the nuclear norm and the spectral norm of ${\mathbf{V}}_i$, respectively. Note that for any ${\mathbf{V}}_i\in\mathbb{H}^{N+1}$ and ${\mathbf{V}}_i \succeq 0$, the equality constraint in~\eqref{nuclear} is met if and only if the constraint~\eqref{rank V} is met.\\
\indent Next, we obtain the following optimization problem by exploiting the penalty method~\cite{ben1997penalty} and incorporating the reformulated rank-one constraint~\eqref{nuclear} into the objective function
\begin{subequations}\label{P6}
\begin{align}
\label{P6O} \mathop {\min }\limits_{ {{\mathbf{V}},{\bm{\rho}}} } \;\;&\sum\nolimits_{ \forall i \in \{t,r\}}\left(\tr\left({\mathbf{V}}_i\bar{\mathbf{F}}_i\right)+\mu\left(\left\|{\mathbf{V}}_i\right\|_*-\left\|{\mathbf{V}}_i\right\|_2\right)\right)\\[-0.1cm]
{\rm{s.t.}}\;\;& \eqref{amplitude1}\text{-}\eqref{auxiliary},
\end{align}
\end{subequations}
where $\mu>0$ denotes the penalty factor. \eqref{P6} is non-convex since its objective function is non-convex.
\begin{theorem}\label{theorem1}
Let ${\mathbf{V}}^q\buildrel \Delta \over =\{{\mathbf{V}}_t^q,{\mathbf{V}}_r^q\}$ denote the optimal solution of~\eqref{P6} with penalty factor $\mu^q$. When $\mu^q$ is sufficiently large, every limit point $\overline{\mathbf{V}}\buildrel \Delta \over=\{\overline{\mathbf{V}}_t, \overline{\mathbf{V}}_r\}$ of the sequence $\{{\mathbf{V}}^q\}$ is an optimal solution of~\eqref{P5}.
\begin{proof}
See Appendix A.
\end{proof}
\end{theorem}

Then, we adopt the SCA technique to solve~\eqref{P6} iteratively with the first-order Taylor expansion~\cite{dinh2010local}. For any feasible point $\mathbf{V}^{\left(n\right)}\buildrel \Delta \over=\{\mathbf{V}^{\left(n\right)}_t,\mathbf{V}^{\left(n\right)}_r\}$ in the $n$-th SCA iteration, a convex upper bound of the term $\left\|{\mathbf{V}}_i\right\|_*-\left\|{\mathbf{V}}_i\right\|_2$ is given by 
\begin{equation}\label{SCA}
\begin{aligned}
f_{\text{SCA}}\left({\mathbf{V}}_i,\mathbf{V}^{\left(n\right)}_i\right)&=
\left\|{\mathbf{V}}_i\right\|_*-\|\mathbf{V}^{\left(n\right)}_i\|_2-{\mathbf{d}_{\text{max}}^{(n)}}^\mathrm{H}\left(\mathbf{V}_i-\mathbf{V}^{\left(n\right)}_i\right)\mathbf{d}_{\text{max}}^{(n)}\\[-0.1cm]
&\ge\left\|{\mathbf{V}}_i\right\|_*-\left\|{\mathbf{V}}_i\right\|_2, \forall i \in \{t,r\},
\end{aligned}
\end{equation}
where $\mathbf{d}_{\text{max}}^{(n)}$ denotes the eigenvector associated with the largest eigenvalue of $\mathbf{V}^{\left(n\right)}_k$.
For the $n$-th SCA iteration, a convex semidefinite program (SDP) problem can be obtained by substituting~\eqref{SCA} into~\eqref{P6}
\begin{subequations}\label{P7}
\begin{align}
\label{OBP7}\mathop {\min }\limits_{ {{\mathbf{V}},{\bm{\rho}}} } \;\;&\sum\nolimits_{ \forall i \in \{t,r\}}\tr\left({\mathbf{V}}_i\bar{\mathbf{F}}_i\right)+\mu \sum\nolimits_{ \forall i \in \{t,r\}} f_{\text{SCA}}\left({\mathbf{V}}_i,\mathbf{V}^{\left(n\right)}_i\right)\\
{\rm{s.t.}}\;\;& \eqref{amplitude1}\text{-}\eqref{auxiliary}.
\end{align}
\end{subequations}

The propose PEN algorithm for ES is summarized in \textbf{Algorithm 1}. In the inner loop, $\mathbf{V}$ is optimized by iteratively solving the SCA approximation~\eqref{P7} for the given penalty factor $\mu$. 
Note that the function value of~\eqref{P6} is upper bounded by the minimum of~\eqref{P7} and the objective function value of~\eqref{P7} is non-increasing in each iteration of the inner loop. The inner loop iteration is guaranteed to converge to a stationary point of~\eqref{P6} with the corresponding $\mu$ \cite{dinh2010local}. In the outer loop, the penalty factor $\mu$ gradually increases as in Step 11 with $\varpi\gg 1$ untill the rank-one constraint violation, \ie, $\max\left\{\left\|{\mathbf{V}}_k\right\|_*-\left\|{\mathbf{V}}_k\right\|_2, \forall k \in\{t,r\}\right\}$, is less than $\epsilon_{\text{p}}$. According to \textbf{Theorem 1}, the optimal solution of~\eqref{P6} with sufficiently large $\mu$ is also the optimal solution of~\eqref{P5}. Furthermore, by discarding auxiliary variables, the solution of~\eqref{P4} can be obtained.\\
\indent The main complexity of Algorithm~\ref{algorithm1} is caused by solving the SDP problem~\eqref{P7} in the inner loop in Step 6, whose computational complexity is approximately $\mathcal{O}\left(N^{3}\right)$ if the interior point method is employed~\cite{bomze2010interior}. Thus, the computational complexity of Algorithm~\ref{algorithm1} is given by $\mathcal{O}\left({I}_\text{out} {I}_\text{in} N^3\right)$, where ${I}_\text{out}$ and ${I}_\text{in}$ are the numbers of iterations for the outer loop and the inner loop, respectively.
\begin{algorithm}[htbp]
\caption{Proposed PEN Algorithm for ES}\label{algorithm1}
\begin{algorithmic}[1]
\STATE {{{\bf Input:} $\bar{\mathbf{F}}_i, \forall\{t,r\}$.}}
\STATE {Initialize feasible point $\mathbf{V}^{(n)}$ with $n=0$, the penalty factor $\mu$.}
\STATE {\bf repeat: outer loop}
\STATE \quad Set inner loop index $n = 0$.
\STATE \quad {\bf repeat: inner loop}
\STATE \quad\quad Get an intermediate solution by solving~\eqref{P7} for given $\mathbf{V}^{(n)}$.
\STATE \quad\quad Update $\mathbf{V}^{(n+1)}$ with the intermediate solution and set $n=n+1$.
\STATE \quad {\bf until} the fractional decrease of~\eqref{OBP7} is less than threshold $\epsilon_{\text{SCA}}$.
\STATE \quad Get the solution of~\eqref{P6} with penalty factor $\mu$.
\STATE \quad Update $\mathbf{V}^{(0)}$ with current solution $\mathbf{V}^{(n)}$.
\STATE \quad Update the penalty factor as $\mu=\omega\mu$.
\STATE {\bf until} the constraint violation is less than threshold $\epsilon_{\text{p}} >0$.
\end{algorithmic}
\end{algorithm}
\subsubsection{Extended penalty-based algorithm for MS}
Compared with the STAR-RIS under the ES protocol, the optimization problem~\eqref{P5} for the STAR-RIS under the MS protocol involves an additional binary constraint~\eqref{MS}. We now focus on dealing with this binary constraint; other constraints are handled similarly to the ES case.\\
\indent First, we relax the binary constraint \eqref{MS} as follows:
\begin{equation}\label{binary}
\begin{aligned}
{\rho}_n^i-\left({\rho}_n^i\right)^2 \ge 0, \forall i\in\{t,r\}, n\in\mathcal{N},
\end{aligned}
\end{equation} 
which always holds for ${\rho}_n^i \in\left[0,1\right]$. Equality holds only if ${\rho}_n^i \in\left\{0,1\right\}$. This constraint is handled by exploiting the penalty method similar to constraint \eqref{rank V} leading to the following optimization problem by incorporating the reformulated constraints \eqref{nuclear} and \eqref{binary} into the objective function
\begin{subequations}\label{P8}
\begin{align}
\mathop {\min }\limits_{ {{\mathbf{V}},{\bm{\rho}}} } \;\;&\sum_{ \forall i \in \{t,r\}}\tr\left({\mathbf{V}}_i\bar{\mathbf{F}}_i\right)+\mu\sum_{ \forall i \in \{t,r\}}\left(\left\|{\mathbf{V}}_i\right\|_*-\left\|{\mathbf{V}}_i\right\|_2\right)+\chi\sum_{ \forall i \in \{t,r\}}\sum_{ \forall n \in \mathcal{N}}\left({\rho}_n^i-\left({\rho}_n^i\right)^2\right)\\[-0.2cm]
{\rm{s.t.}}\;\;& \eqref{amplitude1}\text{-}\eqref{auxiliary}.
\end{align}
\end{subequations}
Here $\chi>0$ denotes the penalty factor. \eqref{P8} is non-convex since its objective function is non-convex.\\
\indent We adopt the SCA technique to solve~\eqref{P8} iteratively with the first-order Taylor expansion. For any feasible point $\left\{{{{\rho}}_n^t}^{(n)},{{{\rho}}_n^r}^{(n)}\right\}$ in the $n$-th SCA iteration, a convex upper bound of the term $\rho_n^i-\left(\rho_n^i\right)^2$ is 
\begin{equation}\label{binarysca}
\begin{aligned}
g_{\text{SCA}}\left(\rho_n^i,{\rho_n^i}^{(n)}\right)=\left({1}-2{\rho_n^i}^{(n)}\right)\rho_n^i+\left({\rho_n^i}^{(n)}\right)^2\ge\rho_n^i-\left(\rho_n^i\right)^2, \forall i \in \{t,r\}, \forall n \in \mathcal{N}.
\end{aligned}
\end{equation}
For the $n$-th SCA iteration, a convex SDP problem can be obtained by substituting~\eqref{SCA} and~\eqref{binarysca} into~\eqref{P8}
\begin{subequations}\label{P9}
\begin{align}
\label{OBP9}\mathop {\min }\limits_{ {{\mathbf{V}},{\bm{\rho}}} } \;\;&\sum_{ \forall i \in \{t,r\}}\tr\left({\mathbf{V}}_i\bar{\mathbf{F}}_i\right)+\mu \sum\nolimits_{ \forall i \in \{t,r\}} f_{\text{SCA}}\left({\mathbf{V}}_i,\mathbf{V}^{\left(n\right)}_i\right)+\chi\sum_{ \forall i \in \{t,r\}}\sum_{ \forall n \in \mathcal{N}}g_{\text{SCA}}\left(\rho_n^i,{\rho_n^i}^{(n)}\right)\\
{\rm{s.t.}}\;\;& \eqref{amplitude1}\text{-}\eqref{auxiliary}.
\end{align}
\end{subequations}

The proposed PEN algorithm for MS is summarized in \textbf{Algorithm 2}. Both the convergence analysis and the computational complexity analysis are similar to that of Algorithm~\ref{algorithm1}.
\begin{algorithm}[!htbp]
\caption{Proposed PEN Algorithm for MS}\label{algorithm2}
\begin{algorithmic}[1]
\STATE {{{\bf Input:} $\bar{\mathbf{F}}_i, \forall\{t,r\}$.}}
\STATE {Initialize feasible points $\mathbf{V}^{(n)}$ and $\bm{\rho}^{(n)}$ with $n=0$, the penalty factor $\mu$ and $\chi$.}
\STATE {\bf repeat: outer loop}
\STATE \quad Set inner loop index $n = 0$ .
\STATE \quad {\bf repeat: inner loop}
\STATE \quad\quad Get an intermediate solution by solving~\eqref{P9} for given $\mathbf{V}^{(n)}$ and $\bm{\rho}^{(n)}$.
\STATE \quad\quad Update $\mathbf{V}^{(n+1)}$ and $\bm{\rho}^{(n+1)}$ with the intermediate solution and set $n=n+1$.
\STATE \quad {\bf until} the fractional decrease of~\eqref{OBP9} is less than threshold $\epsilon_{\text{SCA}}$.
\STATE \quad Get the solution of~\eqref{P8} with penalty factor $\mu$ and $\chi$.
\STATE \quad Update $\mathbf{V}^{(0)}$ and $\bm{\rho}^{(0)}$ with current solution $\mathbf{V}^{(n)}$ and $\bm{\rho}^{(n)}$.
\STATE \quad $\mu=\omega\mu$, $\chi=\varpi\chi$.
\STATE {\bf until} the constraint violation is less than threshold $\epsilon_{\text{p}} >0$.
\end{algorithmic}
\end{algorithm}
\subsection{The Overall Algorithm to Solve~\eqref{P2}}
The proposed algorithm for solving~\eqref{P2} with the STAR-RIS under the ES and MS protocol is summarized in \textbf{Algorithm 3}. The objective function value of~\eqref{P2} is non-decreasing in each iteration of Algorithm~\ref{algorithm3}. Besides, the objective function value of~\eqref{P2} is upper bounded since the transmit power is limited. Hence, Algorithm~\ref{algorithm3} is guaranteed to converge to a stationary point of~\eqref{P2}. The main complexity of Algorithm~\ref{algorithm3} comes from solving~\eqref{P3} and~\eqref{P5} with Algorithm~\ref{algorithm1} or Algorithm~\ref{algorithm2}. The complexity for solving~\eqref{P3} using the Lagrangian multiplier method in \cite{pan2020multicell} is given by $\mathcal{O}\left(K M_b^3\right)$. Thus, the computational complexity of Algorithm~\ref{algorithm3} is given by $\mathcal{O}\left({I}_\text{BCD}\left(K M_b^3 + {I}_\text{out} {I}_\text{in}  N^3\right)\right)$, where ${I}_\text{BCD}$ is the number of BCD iterations.
\begin{algorithm}[htbp]
\caption{Proposed BCD-PEN Algorithm}\label{algorithm3}
\begin{algorithmic}[1]
\STATE {Initialize feasible $\mathbf{\Phi}$ and $\mathbf{W}$ that satisfy~\eqref{Phi} and~\eqref{power}, respectively.}
\STATE {{\bf{repeat:}}}
\STATE {\quad Given $\mathbf{W}$ and $\mathbf{\Phi}$, update the combining matrices $\mathbf{U}$ using \eqref{U}.}
\STATE {\quad Given $\mathbf{W}$, $\mathbf{\Phi}$ and $\mathbf{U}$, update the auxiliary matrices $\mathbf{Z}$ using \eqref{Z}.}
\STATE {\quad Given $\mathbf{\Phi}$, $\mathbf{U}$ and $\mathbf{Z}$, update the transmit beamforming matrices $\mathbf{W}$ by solving~\eqref{P3}.}
\STATE {{\bf{if}} the STAR-RIS works under the ES protocol {\bf{then}}}
\STATE {\quad Given $\mathbf{W}$, $\mathbf{U}$ and $\mathbf{Z}$, update the TRC matrices $\mathbf{\Phi}$ by solving~\eqref{P5} with Algorithm~\ref{algorithm1}.}
\STATE {{\bf{else if}} the STAR-RIS works under the MS protocol {\bf{then}}}
\STATE {\quad Given $\mathbf{W}$, $\mathbf{U}$ and $\mathbf{Z}$, update the TRC matrices $\mathbf{\Phi}$ by solving~\eqref{P5} with Algorithm~\ref{algorithm2}.}
\STATE {{\bf{end if}}}
\STATE  {\bf until} the fractional increase of~\eqref{OBP2} is less than the threshold $\epsilon_{\text{BCD}}$.
\end{algorithmic}
\end{algorithm} 
\section{Low Complexity BCD-ELE Based Solutions}
The optimization of TRCs constitutes a primary source of computational complexity in the BCD-PEN algorithm. In near-field communications, the number of STAR elements is large, \eg, several hundreds of or even thousands of elements. The PEN algorithm becomes impractical to implement due to the extremely high computational complexity. To address this problem, the low-complexity ELE algorithm is proposed to solve~\eqref{P4} in this section. Specifically, the TRCs of each STAR element are optimized in turn with the other $N-1$ elements are fixed. 

Define $f^i_{qj}$ as the element in the $q$-th row and the $j$-th column of matrix $\mathbf{F}_i\in\mathbb{C}^{N\times N}$ and let $e_n^i$ and $v_n^i$ denote the $n$-th element of the vector $\mathbf{e}_i\in\mathbb{C}^{N\times 1}$ and $\mathbf{v}_i\in\mathbb{C}^{N\times 1}$, respectively. Recall that $\mathbf{v}_{i}=\diag\left(\mathbf{\Phi}_{i}\right)$. The term $\mathbf{e}_i^\mathrm{T}\mathbf{v}_i, \forall i\in\{t,r\}$ in \eqref{Gk} can be written as
\begin{equation}\label{e_expand}
\begin{aligned}
\mathbf{e}_i^\mathrm{T}\mathbf{v}_i=\left[e_1^{i},  e_2^{i},  \cdots,  e_N^{i}\right]\left[v_1^{i}, v_2^{i}, \cdots, v_N^{i}\right]^\mathrm{T}= e_n^i v_n^{i} + \sum\nolimits_{j\neq n}^N e_{j}^i v_j^{i},
\end{aligned}
\end{equation}
and $\mathbf{v}_i^\mathrm{H}\mathbf{F}_i\mathbf{v}_i, \forall i\in\{t,r\}$ in \eqref{Gk} can be written as
\begin{equation}\label{F_expand}
\begin{aligned}
&\mathbf{v}_i^\mathrm{H}\mathbf{F}_i\mathbf{v}_i=\left[(v_1^{i})^*  (v_2^{i})^*  \cdots  (v_N^{i})^*\right]
\begin{bmatrix}
f_{1,1}^i & f_{1,2}^i & \cdots & f_{1,N}^i\\
f_{2,1}^i & f_{2,2}^i & \cdots & f_{2,N}^i\\
\vdots & \vdots & \ddots & \vdots\\
f_{N,1}^i & f_{N,2}^i & \cdots & f_{N,N}^i
\end{bmatrix}
\begin{bmatrix}
v_1^{i}\\
v_2^{i}\\
\vdots\\
v_N^{i}
\end{bmatrix}\\
&=\left|v_n^{i}\right|^2 f_{n,n}^i + 2\Re\left(\sum_{q\neq n}^N (v_q^{i})^* f_{q,n}^i v_n^{i}\right) + \sum_{q\neq n}^N\sum_{j\neq n}^N (v_q^{i})^* f_{q,j}^iv_j^{i}.
\end{aligned}
\end{equation}

Substituting \eqref{e_expand} and \eqref{F_expand} into \eqref{Gk}, $\sum_{k\in\mathcal{K}_i}g_k\left(\mathbf{v}\right)$ can be expressed as
\begin{equation}\label{Gk_expand}
\begin{aligned}
\sum_{k\in\mathcal{K}_i}g_k\left(\mathbf{v}\right)&=\left|v_n^{i}\right|^2 f_{n,n}^i+ 2\Re\left(\sum_{q\neq n}^N (v_q^{i})^* f_{q,n}^i v_n^{i}- e_n^iv_n^{i}\right) \\
&\quad+ \sum_{q\neq n}^N\sum_{j\neq n}^N (v_q^{i})^* f_{q,j}^i v_j^{i}-2\Re\left(\sum_{j\neq n}^N e_{j}^iv_j^{i}\right)=\left|v_n^{i}\right|^2A_n^i + 2\Re\left(B_n^iv_n^{i}\right) + C_n^i,
\end{aligned}
\end{equation}
where
\begin{equation}
\begin{aligned}
&A_n^{i}=f_{n,n}^i, \quad B_n^i=\sum_{q\neq n}^N (v_q^{i})^* f_{q,n}^i-e_n^i, \quad C_n^{i}=\sum_{q\neq n}^N\sum_{j\neq n}^N (v_q^{i})^* f_{q,j}^i v_j^{i}-2\Re\left(\sum_{j\neq n}^N e_{j}^iv_j^{i}\right).
\end{aligned}
\end{equation}

For given TRCs of $N-1$ STAR elements, the design of the $n$-th transmission and reflection coefficients in~\eqref{P4}, \ie, $v_n^i=\sqrt{\rho_n^i}e^{j\theta_n^i}, \forall n \in \mathcal{N}, \forall i \in\{t,r\}$, can be formulated as the following problem 
\begin{subequations}\label{P10}
\begin{align}
\mathop {\min }\limits_{ {{v_n^t},{v_n^r}} } \;\;&\sum\nolimits_{ \forall i \in \{t,r\}}\left|v_n^{i}\right|^2A_n^i + 2\Re\left(B_n^iv_n^{i}\right)\\
{\rm{s.t.}}\;\;&{\theta}_n^i, {\rho}_n^i\in \mathcal{C}_{\mathcal{X}},\forall i \in \left\{ {t,r} \right\}.
\end{align}
\end{subequations}

First, we focus on the optimization of phases of the $n$-th STAR element, leading to the following optimization problem 
\begin{subequations}\label{P11}
\begin{align}
\mathop {\min }\limits_{ {{\theta_n^t},{\theta_n^r}} } \;\;&\sum\nolimits_{\forall i \in \{t,r\}}\Re\left(\bar{B}_n^ie^{j\theta_n^i}\right)\\[-0.1cm]
{\rm{s.t.}}\;\;& \theta_n^i \in [0,2\pi), \forall i \in \{t,r\},
\end{align}
\end{subequations}
where $\bar{B}_n^i=\sqrt{\rho_n^i}B_n^i$. Then, the optimal phases of the $n$-th STAR element, \ie, $\theta_n^i, \forall i \in \{t,r\}$ are given by 
\begin{equation}\label{phaseopt}
\begin{aligned}
\theta_n^i = \left\{ {\begin{array}{*{20}{l}}
{\pi  - \angle B_n^i, B_n^i \in[0,\pi)}\\[-0.2cm]
{3\pi  - \angle B_n^i, B_n^i \in[\pi,2\pi)}
\end{array}} \right..
\end{aligned}
\end{equation}

Substituting~\eqref{phaseopt} into~\eqref{P10}, the optimization problem for the amplitudes of the $n$-th STAR element are given by 
\begin{subequations}\label{P12}
\begin{align}
\mathop {\min }\limits_{ {{\rho_n^t},{\rho_n^r}} } \;\;&\sum\nolimits_{ \forall i \in \{t,r\}}\rho_n^iA_n^i - 2\sqrt{\rho_n^i}\left|B_n^i\right|\\[-0.1cm]
{\rm{s.t.}}\;\;& \rho_n^t+\rho_n^r=1,\\[-0.1cm]
\label{amplitude}&\rho_n^t,\rho_n^r\in \{0, 1\}, 
\end{align}
\end{subequations}
where the constraint \eqref{amplitude} is only present when the STAR-RIS works under the MS protocol.
\subsection{Element-wise Iterative Algorithm for ES}
In the following, we first solve~\eqref{P12} for the STAR-RIS under the ES protocol, when constraint~\eqref{amplitude} is absent. Using $\rho_n^r=1-\rho_n^t$,~\eqref{P12} can be reformulated as an unconstrained optimization problem 
\begin{subequations}\label{P13}
\begin{align}
\mathop {\min }_{ {\rho_n^t} } \;\;&\left(A_n^t-A_n^r\right)\rho_n^t-2\sqrt{\rho_n^t}\left|B_n^t\right|-2\sqrt{1-\rho_n^t}\left|B_n^r\right|.
\end{align}
\end{subequations}

It can be easily verified that~\eqref{P13} is convex with respect to $\rho_n^t$. Then, the solution to~\eqref{P13} (as well as the solution to~\eqref{P12}) can be obtained by solving ${{d {f_{n}}} \mathord{\left/
{\vphantom {{d {f_{n}}} {d {\rho_n^t}}}} \right.\kern-\nulldelimiterspace} {d {\rho_n^t}}}=0$, where $f_{n}\left(\rho_n^t\right)$ denotes the objective function of~\eqref{P13}. The derivative of the objective function can be expressed as 
\begin{equation}\label{derivative}
\begin{aligned}
f_{n}'\left(\rho_n^t\right)=\left(A_n^t-A_n^r\right)-\left|B_n^t\right|\frac{1}{\sqrt{\rho_n^t}}+\left|B_n^r\right|\frac{1}{\sqrt{1-\rho_n^t}}.
\end{aligned}
\end{equation}

Note that $f_{n}'\left(\rho_n^t\right)$ is a monotonically increasing function for $\rho_n^t$ and satisfies
\begin{equation}
\begin{aligned}
\mathop {\lim }\limits_{\rho_n^t \to 0^+}f_{n}'\left(\rho_n^t\right) \to -\infty, \mathop {\lim }\limits_{\rho_n^t \to 1^-}f_{n}'\left(\rho_n^t\right) \to \infty. 
\end{aligned}
\end{equation}
The solution of ${{d {f_{n}}} \mathord{\left/{\vphantom {{d {f_{n}}} {d {\rho_n^t}}}} \right.\kern-\nulldelimiterspace} {d {\rho_n^t}}}=0$, \ie, the optimal solution of \eqref{P12}, can be efficiently obtained by using the bisection search algorithm, which is summarized in \textbf{Algorithm 4}. The number of iterations for Algorithm~\ref{algorithm5} to converge is given by $\log_2\left(\frac{1}{\epsilon}\right)$, where $\epsilon$ is the convergence threshold of Algorithm 4. The computational complexity for updating TRCs for the STAR-RIS linearly increases with the number of STAR elements, i.e., $N$. Finally, we have the following proposition:
\begin{proposition}\label{proposition1}
\emph{When the STAR-RIS works under the ES protocol, the optimal solution of }~\eqref{P10} \emph{is given by}
\begin{equation}\label{ESphase}
\begin{aligned}
\theta _n^i = \left\{ {\begin{array}{*{20}{l}}
{\pi  - \angle B_n^i, B_n^i \in[0,\pi)}\\
{3\pi  - \angle B_n^i, B_n^i \in[\pi,2\pi)}
\end{array}} \right., \left\{ {\begin{array}{*{20}{l}}
{\rho _n^t = \left(\rho _n^t\right)^{\text{opt}}}\\
{\rho _n^r = 1-\rho _n^t}
\end{array}} \right.,
\end{aligned}
\end{equation}
\emph{where} $\left(\rho _n^t\right)^{\text{opt}}$ \emph{is the solution of}~\eqref{P12} \emph{obtained by Algorithm 4.}
\end{proposition}
\begin{algorithm}[htbp]
\caption{Bisection Search Algorithm for~\eqref{P13}}\label{algorithm5}
\begin{algorithmic}[1]
\STATE {Initialize feasible $\epsilon>0$, the bounds $\rho^l=0$ and $\rho^u=1$ of $\rho$.}
\STATE {{\bf{repeat:}}}
\STATE {\quad Let $\rho=\left(\rho^l+\rho^u\right)/2$, calculate $f_{n}'\left(\rho\right)$ according to \eqref{derivative}.}
\STATE {\quad{\bf{if}} $f_{n}'\left(\rho\right)>0$}
\STATE {\quad\quad Set $\rho^u=\rho$.}
\STATE {\quad\bf{else}}
\STATE {\quad\quad Set $\rho^l=\rho$.}
\STATE {\quad{\bf{end if}}}
\STATE  {{\bf until} $\left|\rho^l-\rho^u\right|<\epsilon$.}
\STATE {$\rho_n^t=\rho$.}
\end{algorithmic}
\end{algorithm} 
\subsection{Extended Element-wise Iterative Algorithm for MS}
When the STAR-RIS works under the MS protocol, the~\eqref{P12} can be solved by exhaustive search algorithm. The optimal amplitudes of the $n$-th STAR element, \ie, $\rho_n^i, \forall i \in \{t,r\}$ are given by
\begin{equation}\label{amplitudeopt}
\begin{aligned}
\left\{ {\begin{array}{*{20}{c}}
{\left( {\rho _n^t,\rho _n^t} \right) = \left( {0,1} \right),\;{\rm{when}}\;A_n^r - A_n^t - 2\left( {\left| {B_n^r} \right| - \left| {B_n^t} \right|} \right) > 0}\\
{\left( {\rho _n^t,\rho _n^t} \right) = \left( {1,0} \right),\;{\rm{when}}\;A_n^r - A_n^t - 2\left( {\left| {B_n^r} \right| - \left| {B_n^t} \right|} \right) \le 0}
\end{array}} \right..
\end{aligned}
\end{equation} 

Similar to the ES case, the computational complexity for updating TRCs for the STAR-RIS linearly increases with the number of STAR elements. Combining \eqref{phaseopt} and \eqref{amplitudeopt}, we have the following proposition:
\begin{proposition}\label{proposition2}
\emph{When the STAR-RIS works under the MS protocol, the optimal solution of}~\eqref{P10} \emph{is given by}
\begin{equation}\label{MSphase}
\begin{aligned}
\theta _n^i = \left\{ {\begin{array}{*{20}{l}}
{\pi  - \angle B_n^i, B_n^i \in[0,\pi)}\\
{3\pi  - \angle B_n^i, B_n^i \in[\pi,2\pi)}
\end{array}} \right., \left\{ {\begin{array}{*{20}{l}}
{\rho _n^t = u\left(A_n^r-A_n^t-2\left(\left|B_n^r\right|-\left|B_n^t\right|\right)\right)}\\
{\rho _n^r = 1-\rho _n^t}
\end{array}} \right.,
\end{aligned}
\end{equation}
\emph{where} $u\left(\cdot\right)$ \emph{is the Heaviside function.}\vspace{-0.2cm}
\end{proposition}
\subsection{The Overall Algorithm to Solve~\eqref{P2}}
\begin{algorithm}[htbp]
\caption{Proposed Low-Complexity BCD-ELE Algorithm}\label{algorithm4}
\begin{algorithmic}[1]
\STATE {Initialize feasible $\mathbf{\Phi}$ and $\mathbf{W}$ that satisfy~\eqref{Phi} and~\eqref{power}, respectively.}
\STATE {{\bf{repeat:}}}
\STATE {\quad Given $\mathbf{W}$ and $\mathbf{\Phi}$, update the combining matrices $\mathbf{U}$ using~\eqref{U}.}
\STATE {\quad Given $\mathbf{W}$, $\mathbf{\Phi}$ and $\mathbf{U}$, update the auxiliary matrices $\mathbf{Z}$ using~\eqref{Z}.}
\STATE {\quad Given $\mathbf{\Phi}$, $\mathbf{U}$ and $\mathbf{Z}$, update the transmit beamforming matrices $\mathbf{W}$ by solving~\eqref{P3}.}
\STATE {{\bf{for}} $n\in\mathcal{N}$\bf{:}}
\STATE {\quad{\bf{if}} the STAR-RIS works under the ES protocol {\bf{then}}}
\STATE {\quad\quad Given $\mathbf{W}$, $\mathbf{U}$, $\mathbf{Z}$ and TRCs for other $N-1$ STAR elements, update $v_n^t$ and $v_n^r$ using~\eqref{ESphase}.}
\STATE {\quad{\bf{else if}} the STAR-RIS works under the MS protocol {\bf{then}}}
\STATE {\quad\quad Given $\mathbf{W}$, $\mathbf{U}$, $\mathbf{Z}$ and TRCs for other $N-1$ STAR elements, update $v_n^t$ and $v_n^r$ using~\eqref{MSphase}.}
\STATE {\quad{\bf{end if}}}
\STATE {{\bf{end for}}}
\STATE  {\bf until} the fractional increase of \eqref{OBP2} is less than the threshold $\epsilon_{\text{BCD}}$.
\end{algorithmic}
\end{algorithm} 
The proposed low-complexity BCD-ELE algorithm for solving~\eqref{P2} is summarized in~\textbf{Algorithm 5}. {Since the objective function of~\eqref{P10} is minimized in each iteration, both~\textbf{Proposition 1} and~\textbf{Proposition 2} can guarantee to yield a monotonically decreasing objective function value of~\eqref{P2} compared to the previous TRC solution.} Thus, the objective function value of~\eqref{P2} is non-decreasing in each iteration of Algorithm~\ref{algorithm4}. Additionally, since power is limited, the objective function value of~\eqref{P2} has an upper bound. Hence, Algorithm~\ref{algorithm4} is guaranteed to converge to a stationary point of~\eqref{P2}. The main complexity for the Algorithm~\ref{algorithm4} comes from solving the~\eqref{P3} and updating the TRC matrices, at the computational complexity level of $\mathcal{O}\left(K M_b^3\right)$ and $\mathcal{O}\left(N\right)$, respectively. Summarizing all above, the aggregated complexity of Algorithm~\ref{algorithm4} can be expressed as $\mathcal{O}\left({I}_\text{BCD}\left(K M_b^3 +  N\right)\right)$, where ${I}_\text{BCD}$ is the number of iterations needed by Algorithm~\ref{algorithm4}. Compared to the BCD-PEN algorithm with a computational complexity of $\mathcal{O}\left({I}_\text{BCD}\left(K M_b^3 + {I}_\text{out} {I}_\text{in}N^3\right)\right)$, the BCD-ELE algorithm has a significantly lower computational complexity, which will be further verified in the next section. 
\section{Numerical Results}
In this section, numerical results are provided to verify the performance improvement brought by the STAR-RIS as well as the near-field beamforming in MIMO communication systems.
\subsection{Simulation Setup}
As shown in Fig.~\ref{user_setup}, the BS and the STAR-RIS are deployed at $\left({0, 0, 0}\right)$ m and $\left({0, 50, 0}\right)$ m, respectively. All users lie on the circles surrounding the STAR-RIS with a radius of $d_t^1=d_r^1=2$~m or with a radius of $d_t^2=d_r^2=4$~m. The angles of the BS-STAR channel paths are randomly generated following the uniform distribution. Channel path-loss coefficient is modeled as $\beta=C_{0}\left(\frac{d}{D_{0}}\right)^{-\alpha}$, where $C_{0}=-30$ dB is the path loss at reference distance $D_{0}=1$ m. $d$ denotes the distances between the BS and the STAR-RIS. $\alpha=2.2$ denotes path loss exponent of the BS-STAR-RIS channel. Other system parameters are set as follows: $f_c=10$ GHz, $\lambda_c=0.03$ m, $L=16$, $\sigma^{2}=-110$ dBm, $M_b=16$, $M=4$, $K=4$ with $\mathcal{K}_t=\left\{1,2\right\}$ and $\mathcal{K}_r=\left\{3,4\right\}$, weighting factors $ \eta_k=1, \forall k$. The Rayleigh distance of a UPA-type STAR-RIS with $40=5\times 8$ elements operating at $10$ GHz is about $5$ m. This indicates that all users lie in the near-field of the STAR-RIS.\\
\indent Besides, the main adopted simulation parameters for the optimization algorithms are given in Table \ref{simulation_parameters}.
\begingroup
\renewcommand{\arraystretch}{1} 
\begin{table*}[h]
\centering
\begin{footnotesize}
\caption{Simulation Parameters for the Optimization Algorithms}
\label{simulation_parameters}
\centering
\begin{tabular}{|c|c|c||c|c|c|}
\hline
$\mu$, $\chi$&  Penalty factors for Algorithms~\ref{algorithm1} and~\ref{algorithm2}& $10^{-4}$&  $\omega$, $\varpi$& Scaling factors for Algorithms~\ref{algorithm1} and~\ref{algorithm2}& $10$ \\
\hline
$\epsilon_{\text{SCA}}$&  Coverage tolerance for SCA& $10^{-2}$&  $\epsilon_{\text{BCD}}$ & Coverage tolerance for BCD& $10^{-3}$ \\
\hline
$n_\text{in}$, $n_\text{out}$&  Maximum inner/outer iterations& $30$&  $\epsilon_{\text{p}}$& Coverage tolerance for Algorithms~\ref{algorithm1} and~\ref{algorithm2}& $10^{-5}$ \\
\hline
\end{tabular}
\end{footnotesize}
\end{table*}
\subsection{Baseline Schemes and User Setups}
\begin{enumerate}
\item \textbf{Conventional RIS:} Instead of using the STAR-RIS, baseline 1 utilizes a conventional reflecting-only RIS and a transmitting-only RIS, both possessing $N/2$ elements. The optimization problem resulting from this configuration can be resolved by Algorithm~\ref{algorithm3} and~\ref{algorithm4} with given $\bm{\rho}_t=\left[\mathbf{1}_{1\times N/2}\ \mathbf{0}_{1\times N/2}\right]$ and $\bm{\rho}_r=\left[\mathbf{0}_{1\times N/2}\ \mathbf{1}_{1\times N/2}\right]$.
\item \textbf{Uniform energy splitting STAR-RIS:} All the elements of the ES STAR-RIS utilize equal amplitude coefficients for transmission and reflection, respectively. The optimization problem resulting from this configuration can be resolved by Algorithm~\ref{algorithm3} and~\ref{algorithm4} with given $\bm{\rho}_t=\bm{\rho}_r=\frac{1}{2}\mathbf{1}_{1\times N}$.
\item \textbf{Far-field \los channel-based beamforming:} To confirm the benefits brought by the near-field beamforming, this baseline scheme adopts the far-field \los channel model in~\eqref{farH} to represent the STAR-user channels. We set $\beta_k=\left(\alpha^k_{1,1}\right)^2$ to guarantee the fair comparison between the beamforming under far-field channel and near-field channel. The optimization problem resulting from this configuration can be resolved by Algorithm~\ref{algorithm3} and~\ref{algorithm4}.
\end{enumerate}
\begin{figure*}[!htbp]
\centering
\begin{minipage}[t]{0.49\textwidth}
\subfigure[Random user setup.]{\label{Random_user}
\includegraphics[width= 3.2in]{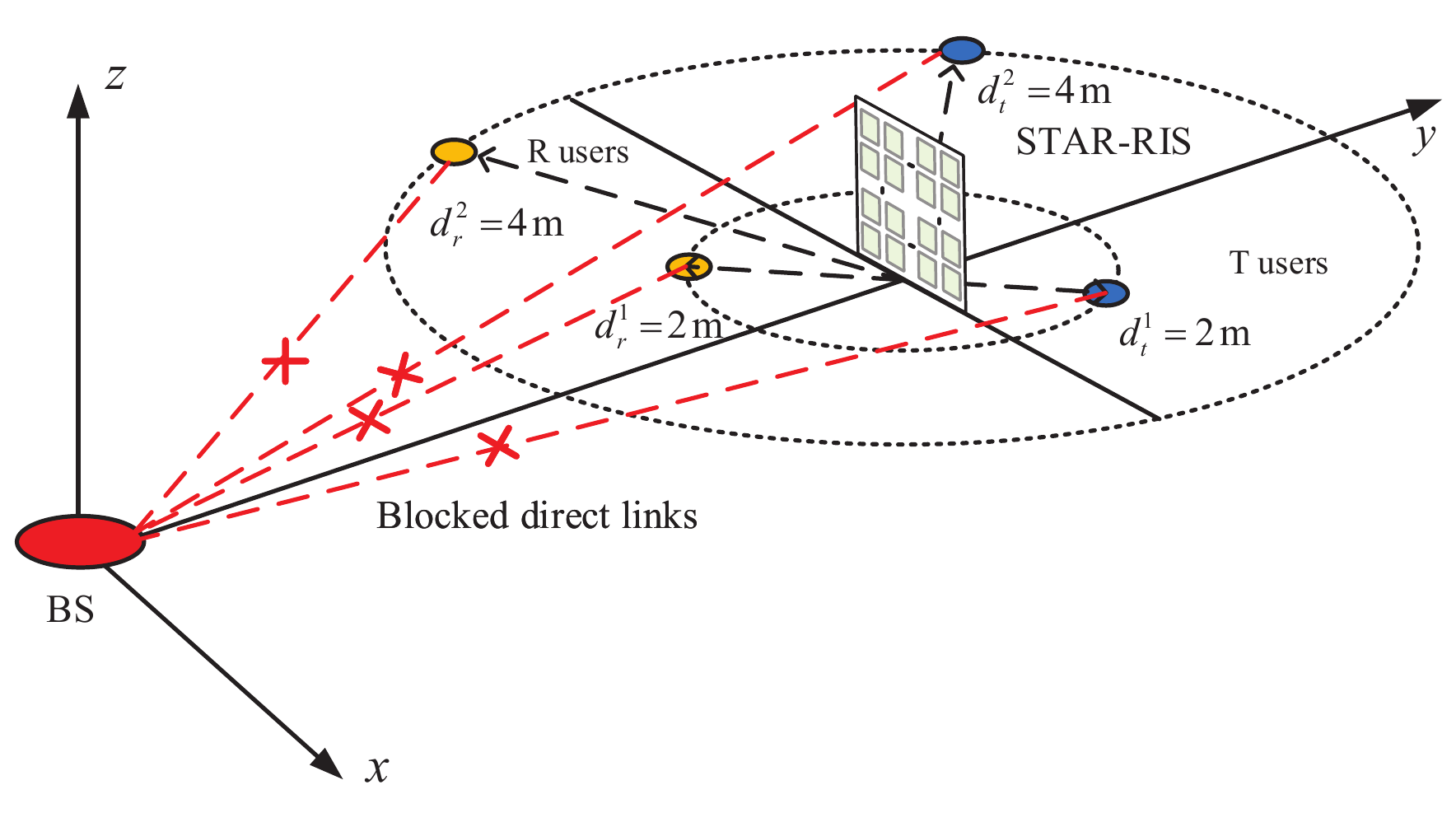}}
\end{minipage}
\begin{minipage}[t]{0.49\textwidth}
\subfigure[Inline user setup.]{\label{Inline_user}
\includegraphics[width= 3.2in]{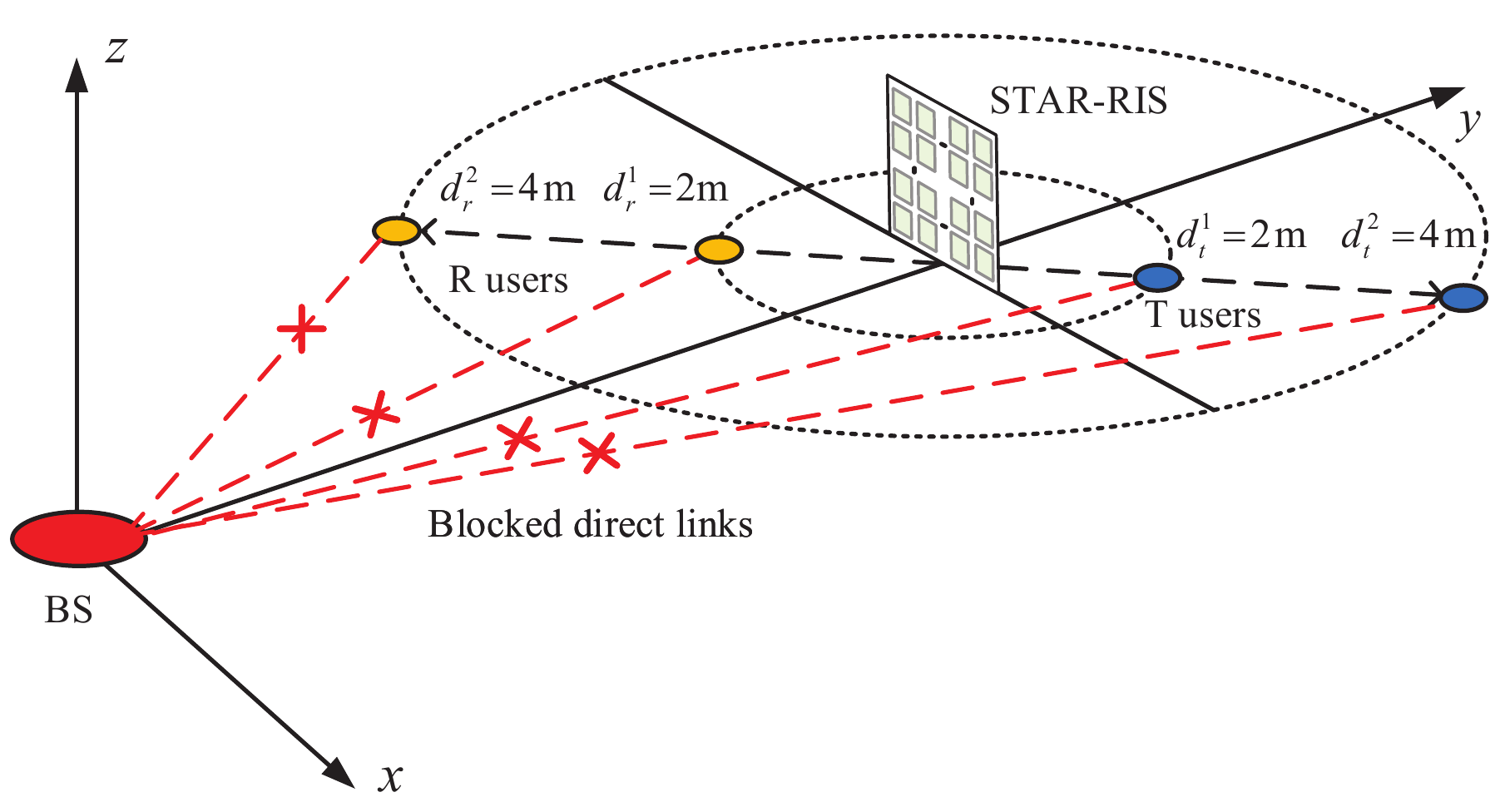}}
\end{minipage}
\caption{Two user setups.}\label{user_setup}
\end{figure*}

To demonstrate the DoFs enhancement and the accessibility improvement brought by the near-field beamforming, we consider two user setups in simulations: 
\begin{enumerate}
\item \textbf{Random user setup:} In this user setup, users in the same region line in different angles with respect to the STAR-RIS, as shown in Fig.~\ref{Random_user}.
\item \textbf{Inline user setup:} In this user set up, all users in the same region line in the same angle with respect to the STAR-RIS, as shown in Fig.~\ref{Inline_user}.
\end{enumerate}

Under the inline user setup, existing works on far-field STAR-RIS aided communications suffer from significant sum rate degradation due to severe inter-user interference~\cite{niu2021weighted,perera2022sum,9935266}. This is because the parallel-wave-based far-field channel only relies on the angle information thus the channels of users in similar angle are highly correlated. The BS is unable to achieve effective inter-user interference management. In contrast, the spherical-wave-based near-field channel contains not only the angle information but also the distance information. The extra distance information helps the BS to focusing signal on intended users and mitigate inter-user interference.
\subsection{Convergence Behavior of BCD Algorithm}
\begin{figure}[!htbp]
\vspace{-0.8cm}
\centering
\begin{minipage}[t]{0.49\textwidth}
\centering
\includegraphics[width=3.4in]{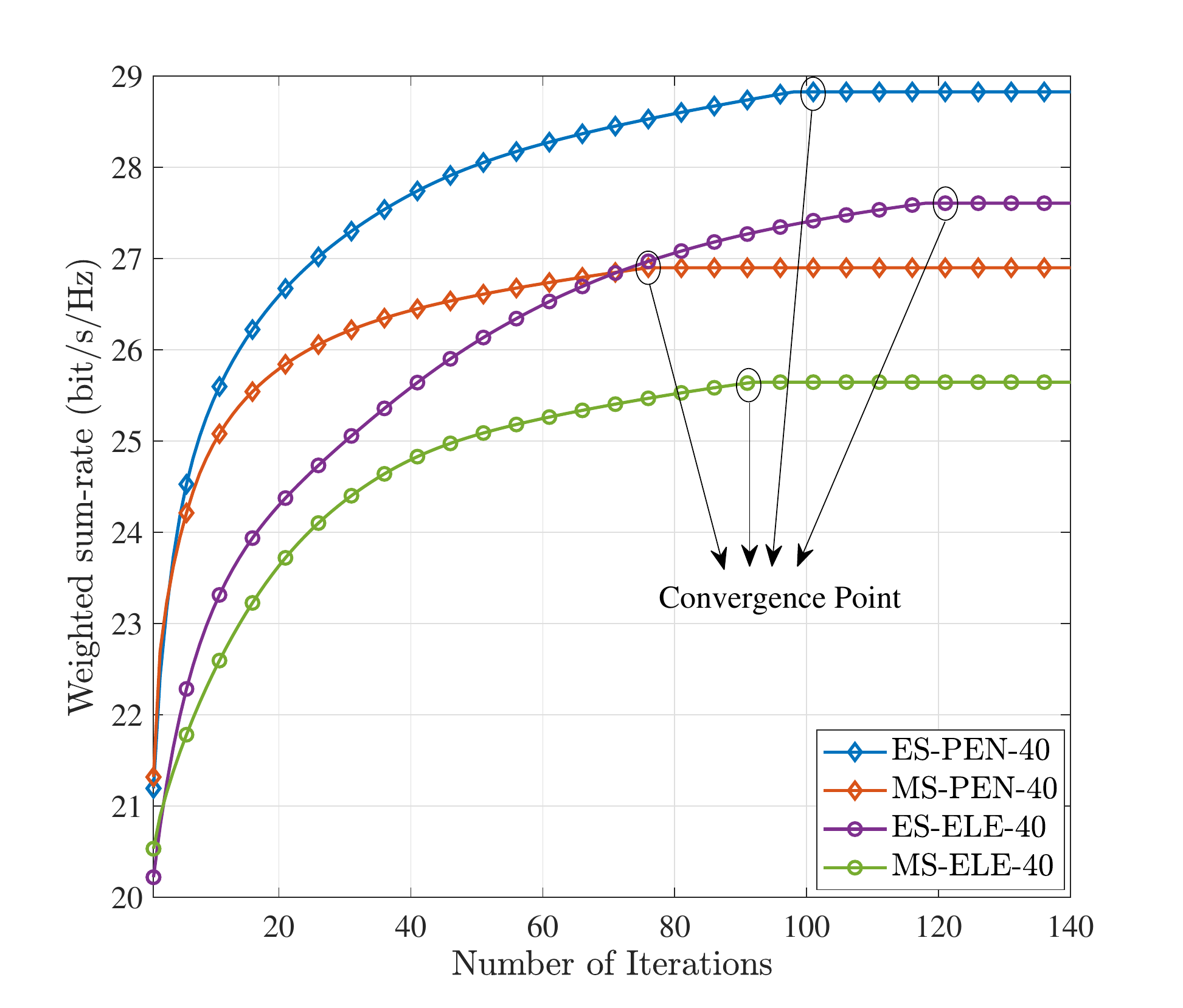}
\caption{Convergence behavior of the proposed algorithms.}\label{Convergence}
\end{minipage}
\begin{minipage}[t]{0.49\textwidth}
\centering
\includegraphics[width=3.4in]{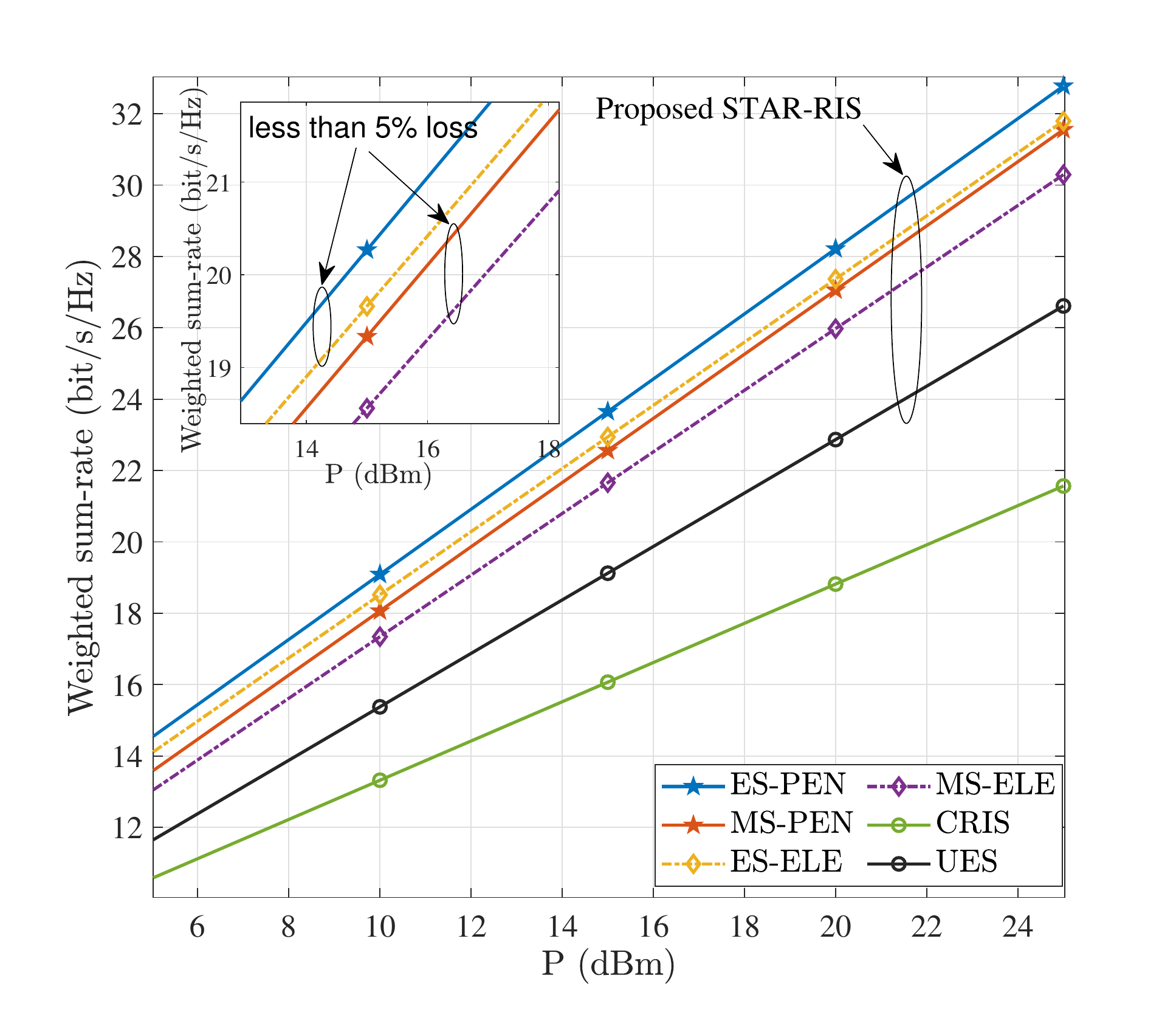}
\caption{Weighted sum rate of users versus the transmit power of the BS.}\label{PEN_ELE_UES_CRIS}
\end{minipage}
\end{figure}

We ﬁrst study the convergence behaviour of the proposed BCD-PEN algorithm and the proposed BCD-ELE algorithm. In Fig.~\ref{Convergence}, we show the convergence speed and the achievable weighted sum rate of the proposed algorithms when $N=40$. The overall complexity and running time for each iteration of the proposed algorithms are provided in Table~\ref{Complexity_and_Running_Time}. All simulations are carried on the same computer with a 2.20GHz Intel(R) Core(TM) i7-8750H CPU and 16GB RAM. For both ES and MS protocols, the proposed algorithms converge within 130 iterations. We can observe from Fig.~\ref{Convergence} that using the ES protocol leads to a slightly slower convergence speed than the MS protocol for both the BCD-PEN and BCD-ELE algorithms. This is due to the fact that more DoFs are involved when the STAR-RIS works under the ES protocol, and more iterations are required for convergence. Moreover, the BCD-ELE algorithm takes more iterations to converge than the BCD-PEN algorithm under both ES and MS protocols. The reason behind this is the BCD-PEN algorithm updates the configuration of the STAR-RIS as a whole in each iteration while the BCD-ELE algorithm updates the configuration of the STAR-RIS in an element-wise manner in each iteration. Therefore, the BCD-ELE algorithm in general requires more iterations to converge than the BCD-PEN algorithm. Nevertheless, it is worth mentioning that the BCD-ELE algorithm is still more efficient than the BCD-PEN algorithm. The reason behind this is that the PEN algorithm used in every iteration of the BCD-PEN algorithm has a significantly larger computation complexity than the ELE algorithm used in every iteration of the BCD-ELE algorithm. From Table~\ref{Complexity_and_Running_Time}, we can observe that each iteration of the BCD-ELE algorithm requires much less running time than the BCD-PEN algorithm. This is consistent with the complexity analysis of the proposed algorithms.
\begingroup
\renewcommand{\arraystretch}{1} 
\begin{table*}[h]
\centering
\begin{footnotesize}
\caption{Complexity and Running Time Comparison for each Iteration of Proposed Algorithms}
\label{Complexity_and_Running_Time}
\centering
\begin{tabular}{|c|c|c|}
\hline
Algorithm & Complexity per iteration& Running time per iteration \\
\hline
the BCD-PEN algorithm for ES &  $\mathcal{O}\left(K M_b^3 + {I}_\text{out} {I}_\text{in}  N^3\right)$ & $91.5511$ second \\
\hline
the BCD-PEN algorithm for MS &  $\mathcal{O}\left(K M_b^3 + {I}_\text{out} {I}_\text{in}  N^3\right)$ & $89.3249$ second \\
\hline
the BCD-ELE algorithm for ES & $\mathcal{O}\left(K M_b^3 + N\right)$ & $0.0402$ second \\
\hline
the BCD-ELE algorithm for MS & $\mathcal{O}\left(K M_b^3 + N\right)$  &  $0.0457$ second \\
\hline
\end{tabular}
\end{footnotesize}
\end{table*}
\subsection{Weighted Sum Rate Versus the Transmit Power}
In Fig.~\ref{PEN_ELE_UES_CRIS}, we investigate the achieved weighed sum rate versus the transmit power. We set $N=40$ and adopt the random user setup in the simulation. As we can see from the figure, the weighted sum rate for all schemes and protocols increases as the BS transmit power increases. This is expected since more power budget at the BS allow the users receive stronger signals. It can be observed that regardless the adopted operating protocol, the STAR-RIS always outperform conventional RIS baseline. Moreover, Fig.~\ref{PEN_ELE_UES_CRIS} also reveals that the low complexity BCD-ELE algorithm achieves almost the same performance as the BCD-PEN algorithm. However, compared with BCD-PEN algorithm, the computation complexity of the BCD-ELE algorithm increases linearly with the number of STAR elements $N$, which makes it more appealing in near-field communication where the STAR-RIS array is large. 
\begin{figure*}[ht]
\centering
\subfigure[Weighted sum rate of users versus the transmit power of the BS under different user setups with $N=40$.]{\label{40_inline_random}
\includegraphics[width= 3.1in]{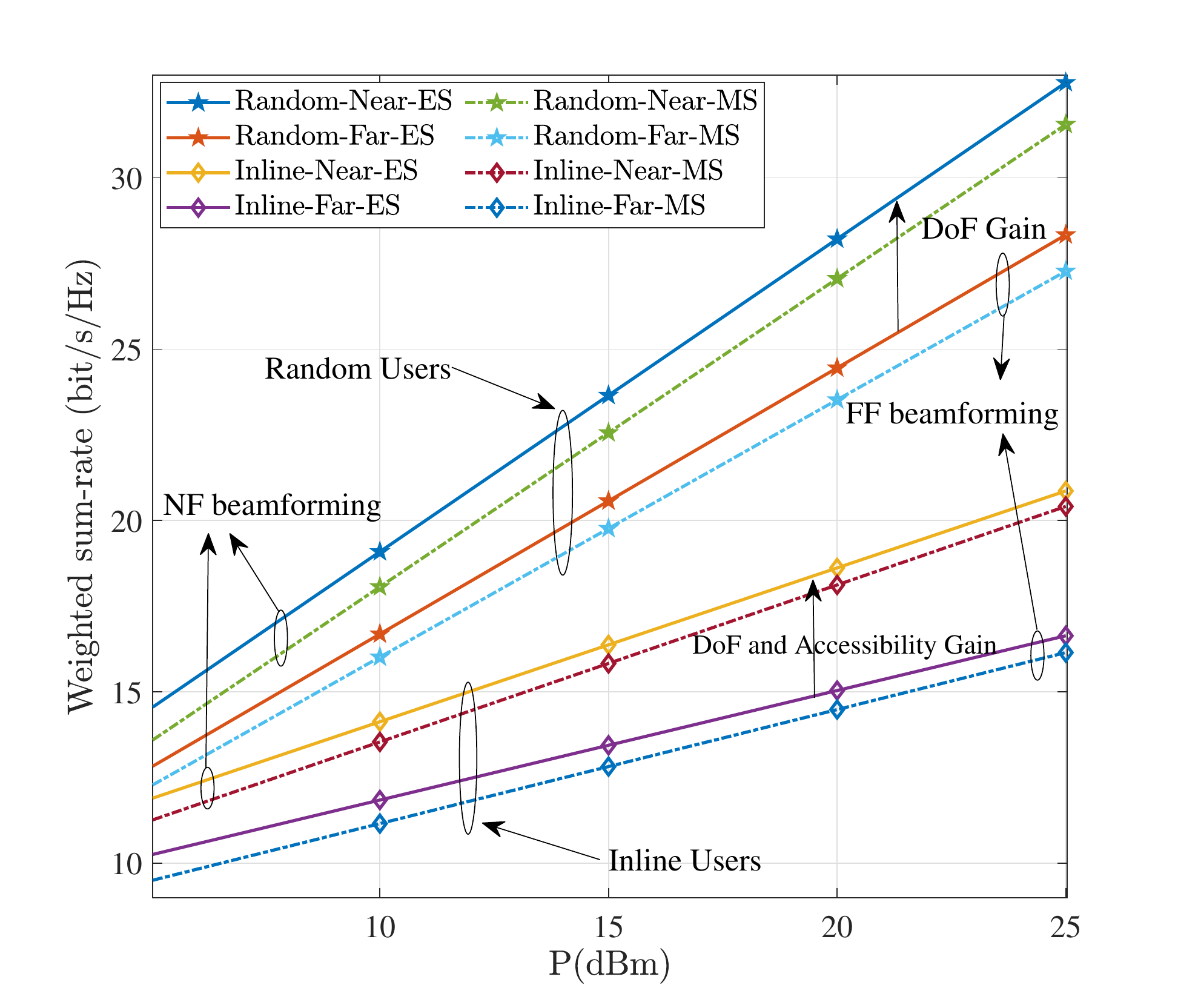}}
\subfigure[Weighted sum rate of users versus the transmit power of the BS under different user setups with $N=400$.]{\label{400_inline_random}
\includegraphics[width= 3.1in]{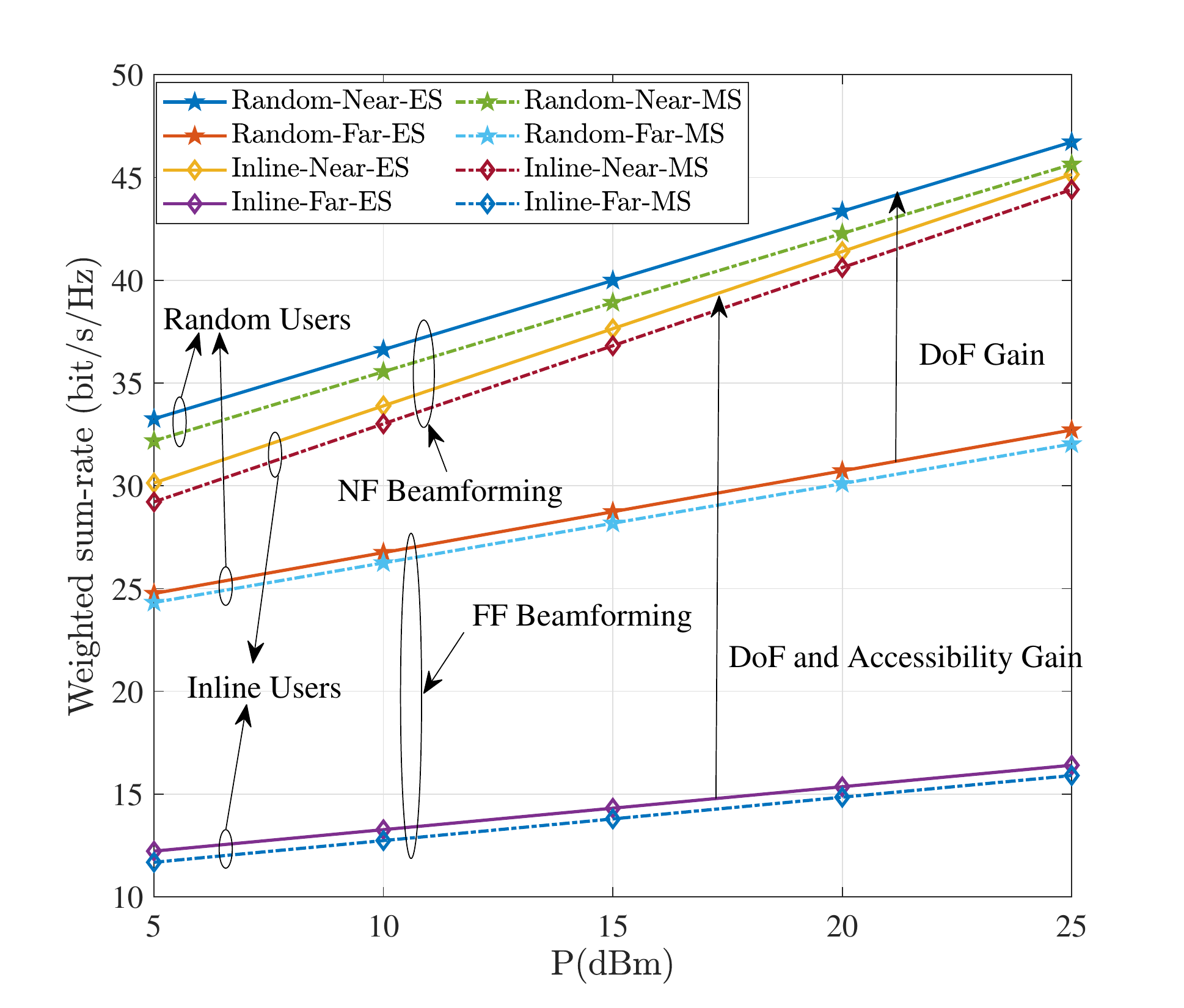}}
\caption{Weighted sum rate of users versus the transmit power of the BS.}\label{inline_random}
\end{figure*}
\subsection{Impact of the Near-field Beamforming}
In Fig.~\ref{inline_random}, we investigate the impact of the near-field beamforming under both the random user setup and the inline user setup. The results of Fig~\ref{40_inline_random} is generated with the BCD-PEN algorithm. The results of Fig~\ref{400_inline_random} is generated with the BCD-ELE algorithm due to the large STAR elements number. As can be seen from Fig.~\ref{40_inline_random} and Fig.~\ref{400_inline_random}, ES outperforms MS for all schemes and user location setups, which is consistent with Fig.~\ref{PEN_ELE_UES_CRIS}. For the same beamforming scheme (far-field or near-field beamforming), it can be observed from Fig.~\ref{40_inline_random} and Fig.~\ref{400_inline_random} that the random user setup always outperforms the inline user setup. This is expected since the inline user setup will lead to high inter-user interference which inevitably leads to sum rate degradation.\\
\indent For the same user setup (random or inline user setup), it can be observed from both Fig.~\ref{40_inline_random} and Fig.~\ref{400_inline_random} that the near-field beamforming always provides higher capacity than the far-field beamforming. Specifically, under the random user setup, the capacity gain mainly comes from the DoFs enhancement of near-field channel, \ie, the near-field \los channels have higher rank and can carry more data streams to the multi-antenna users. Under the inline user setup, the capacity gain comes from both the DoFs enhancement and the accessibility improvement of the near-field channel, \ie, except for the gain brought by the high rank near-field \los channels, the user distance information carried by the near-field channel is leveraged by the joint beamforming algorithm to mitigate the severe inter-user interference. What’s more, compare Fig.~\ref{40_inline_random} and Fig.~\ref{400_inline_random}, we can confirm the DoFs and accessibility gain is more pronounced with a larger STAR-RIS array where the near-field effects are more evident. 
\vspace{-0.3cm}
\subsection{Weighted Sum Rate Versus the number of STAR Elements}
\begin{figure}[!htbp]
\vspace{-0.8cm}
\centering
\includegraphics[width=3.5in]{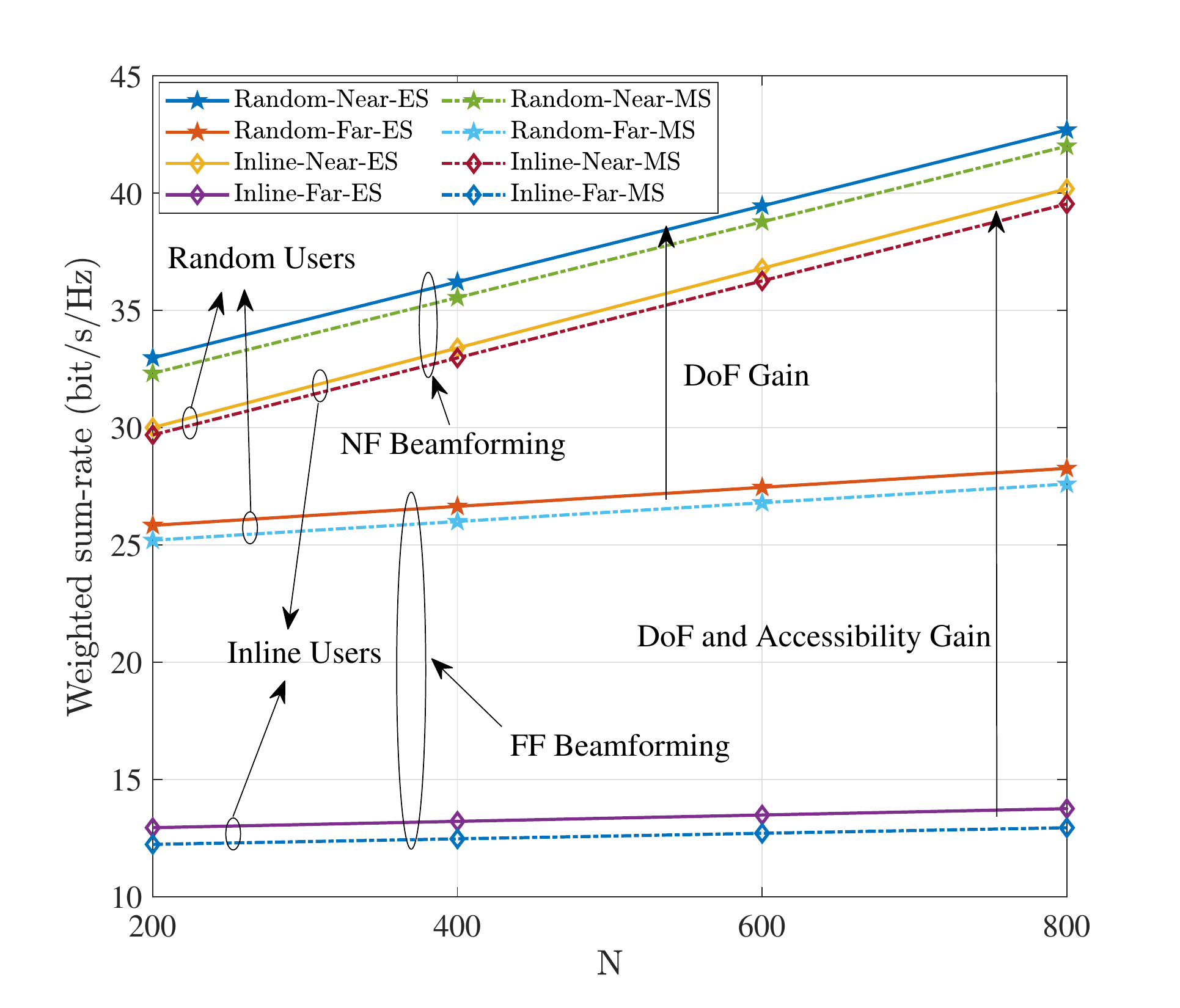}
\caption{Weighted sum rate of users versus the STAR elements number $N$ under different user setups with $P=10$ dBm.}\label{ES-MS-10dBm}
\end{figure}
As can be seen from Fig.~\ref{ES-MS-10dBm}, the weighted sum rate of users under all schemes and user setups increases as the STAR elements number $N$ increases. This is expected since larger $N$ enable a higher transmission/reﬂection beamforming gain. 
For the near-field beamforming, not only the angle information but also the distance information of users is used to help the BS and the STAR-RIS to focusing signal on intended users and mitigate inter-user interference, thus the performance gap between the inline user setup and the random user setup is smaller than that in the far-field beamforming scheme. For the far-field beamforming, especially under inline user setup, the improvement in the weighted sum rate for users brought by the increasing STAR elements number is insignificant. This is because increasing $N$ leads to both higher intended signals and inter-user interference under far-field beamforming, resulting in a limited improvement in the user signal-to-interference-plus-noise ratio (SINR). This limitation causes the performance gap between the far-field beamforming scheme and the near-field beamforming scheme to become more pronounced as $N$ increases. 
\vspace{-0.2cm}
\section{Conclusion}
A STAR-RIS aided near-field MIMO communication framework was proposed. A weighted sum rate maximization problem for the joint optimization of the active beamforming at the BS and the TRCs at the STAR-RIS was formulated. The resulting non-convex problem was first reformulated into an equivalent problem using the WMMSE method. Then, the equivalent problem was solved by the developed BCD-based algorithm. In particular, under given STAR-RIS TRCs, the optimal active beamforming matrices were obtained by solving a convex quadratically constrained quadratic program problem. Under given active beamforming matrices, the PEN algorithm and the ELE algorithm were conceived for optimizing the STAR-RIS TRCs. Numerical results confirmed that i) the near-field beamforming could significantly enhance the weighted sum rate for STAR-RIS aided multi-user MIMO systems; {and ii) the near-field channels facilitated by the STAR-RIS provide enhanced DoFs and accessibility for the multiuser MIMO system; and iii) the BCD-PEN algorithm achieves better performance than the BCD-ELE algorithm, while the latter has a significantly lower computational complexity.}
\numberwithin{equation}{section}
\section*{Appendix~A: Proof of Theorem~\ref{theorem1}} \label{Appendix:A}
\renewcommand{\theequation}{A.\arabic{equation}}
\setcounter{equation}{0}
Let $\widehat{g}\left(\mathbf{V}\right)$ and $\mathbf{V}^\dag\buildrel \Delta \over=\{\mathbf{V}^\dag_t,\mathbf{V}^\dag_r\}$ denote the objective function and the optimal solution of~\eqref{P5}, respectively. For any feasible $\mathbf{V}$, we have\vspace{-0.2cm}
\begin{equation}\label{THEOREM}
\begin{aligned}
\widehat{g}\left(\mathbf{V}\right) \ge \widehat{g}\left(\mathbf{V}^\dag\right).
\end{aligned}\vspace{-0.2cm}
\end{equation}

Let $\dot{g}\left(\mathbf{V};\mu^q\right)$ and $\mathbf{V}^q\buildrel \Delta \over=\{\mathbf{V}^q_t,\mathbf{V}^q_r\}$ denote the objective function of~\eqref{P6} and the optimal solution of~\eqref{P6} with penalty factor $\mu^q$, respectively. Then, we have 
\begin{equation}\label{THEOREM1}
\begin{aligned}
\widehat{g}&\left(\mathbf{V}^q\right)+\mu^q\sum\nolimits_{k\in{t,r}}\left(\left\|{\mathbf{V}}_k^q\right\|_*-\left\|{\mathbf{V}}_k^q\right\|_2\right)=\dot{g}\left(\mathbf{V}^q;\mu^q\right)\\
&\le \widehat{g}\left(\mathbf{V}^\dag\right)+\mu^q\sum\nolimits_{k\in{t,r}}\left(\left\|{\mathbf{V}}_k^\dag\right\|_*-\left\|{\mathbf{V}}_k^\dag\right\|_2\right) =\dot{g}\left(\mathbf{V}^\dag;\mu^q\right)\buildrel (a) \over=\widehat{g}\left(\mathbf{V}^\dag\right),
\end{aligned}
\end{equation}
where $(a)$ comes from the fact that $\left\|{\mathbf{V}}_k^\dag\right\|_*-\left\|{\mathbf{V}}_k^\dag\right\|_2=0$, $\forall k \in\{t,r\}$ since the the optimal solution of~\eqref{P5} must satisfy the constraint \eqref{rank V}.

Since the penalty factor $\mu^q>0$, $\forall q$, from \eqref{THEOREM1} we can get\vspace{-0.1cm}
\begin{equation}\label{THEOREM2}
\begin{aligned}
\sum\nolimits_{k\in{t,r}}\left(\left\|{\mathbf{V}}_k^q\right\|_*-\left\|{\mathbf{V}}_k^q\right\|_2\right) \le \frac{1}{\mu^q}\left(\widehat{g}\left(\mathbf{V}^\dag\right)-\widehat{g}\left(\mathbf{V}^q\right)\right).
\end{aligned}\vspace{-0.1cm}
\end{equation}

As defined in Theorem~\ref{theorem1}, $\overline{\mathbf{V}}$ is a limit point of the sequence $\{\mathbf{V}^q\}$, an infinite subsequence $\mathcal{Q}$ can be found such that\vspace{-0.2cm}
\begin{equation}
\begin{aligned}
\mathop {\lim }\limits_{q \in \mathcal{Q}} {{\bf{V}}} = \overline{\mathbf{V}}.
\end{aligned}\vspace{-0.2cm}
\end{equation}

By taking the limit on both sides of \eqref{THEOREM2}, we can get
\begin{equation}\label{THEOREM3}
\begin{aligned}
\sum\nolimits_{k\in\{t,r\}}&\left(\left\|\overline{\mathbf{V}}_k\right\|_*-\left\|\overline{\mathbf{V}}_k\right\|_2\right)=\mathop {\lim }\limits_{q \in \mathcal{Q}} \frac{1}{\mu^q}\left(\widehat{g}\left(\mathbf{V}^\dag\right)-\widehat{g}\left(\mathbf{V}^q\right)\right)\buildrel (b) \over=0.
\end{aligned}
\end{equation} 
where equality (b) holds for $\mu^q \to \infty$. \eqref{THEOREM3} proves $\overline{\mathbf{V}}$ is a feasible solution of~\eqref{P5}.

Then, by taking the limit on both sides of \eqref{THEOREM1} and using the nonnegativity of the term $\mu^q\sum_{t,r}\left(\left\|{\mathbf{V}}_k^q\right\|_*-\left\|{\mathbf{V}}_k^q\right\|_2\right)$, we have 
\begin{equation}\label{THEOREM4}
\begin{aligned}
\widehat{g}\left(\overline{\mathbf{V}}\right) \le \widehat{g}\left(\mathbf{V}^\dag\right).
\end{aligned}
\end{equation} 

Since $\overline{\mathbf{V}}$ is a feasible solution of~\eqref{P5}, we have following inequality basd on \eqref{THEOREM}
\begin{equation}
\begin{aligned}
\widehat{g}\left(\overline{\mathbf{V}}\right) \ge \widehat{g}\left(\mathbf{V}^\dag\right).
\end{aligned}
\end{equation}

Then, we have 
\begin{equation}
\begin{aligned}
\widehat{g}\left(\overline{\mathbf{V}}\right) = \widehat{g}\left(\mathbf{V}^\dag\right).
\end{aligned}
\end{equation}

Since $\overline{\mathbf{V}}$ is feasible for~\eqref{P5} and its objective value is euqal to that of the optimal solution $\mathbf{V}^\dag$, $\overline{\mathbf{V}}$ is an optimal solution of~\eqref{P5}. This completes the proof.
\bibliography{myref}

\begin{thebibliography}{10}
\providecommand{\url}[1]{#1}
\csname url@samestyle\endcsname
\providecommand{\newblock}{\relax}
\providecommand{\bibinfo}[2]{#2}
\providecommand{\BIBentrySTDinterwordspacing}{\spaceskip=0pt\relax}
\providecommand{\BIBentryALTinterwordstretchfactor}{4}
\providecommand{\BIBentryALTinterwordspacing}{\spaceskip=\fontdimen2\font plus
\BIBentryALTinterwordstretchfactor\fontdimen3\font minus
  \fontdimen4\font\relax}
\providecommand{\BIBforeignlanguage}[2]{{%
\expandafter\ifx\csname l@#1\endcsname\relax
\typeout{** WARNING: IEEEtran.bst: No hyphenation pattern has been}%
\typeout{** loaded for the language `#1'. Using the pattern for}%
\typeout{** the default language instead.}%
\else
\language=\csname l@#1\endcsname
\fi
#2}}
\providecommand{\BIBdecl}{\relax}
\BIBdecl

\bibitem{Li}
H.~Li, Y.~Liu, X.~Mu, Y.~Chen, and Z.~Pan, ``Joint beamforming for {STAR}-{RIS}
  in near-field communications,'' in \emph{Proc. {IEEE} Global Commun. Conf.
  (GLOBECOM)}, Dec. 2023, Submitted.

\bibitem{9349624}
W.~Jiang, B.~Han, M.~A. Habibi, and H.~D. Schotten, ``The road towards 6{G}: A
  comprehensive survey,'' \emph{{IEEE} {O}pen j. Commun. Soc.}, vol.~2, pp.
  334--366, 2021.

\bibitem{9690478}
Y.~Liu, X.~Mu, J.~Xu, R.~Schober, Y.~Hao, H.~V. Poor, and L.~Hanzo, ``{STAR}:
  Simultaneous transmission and reflection for 360° coverage by intelligent
  surfaces,'' \emph{{IEEE} Trans. Wireless Commun.}, vol.~28, no.~6, pp.
  102--109, 2021.

\bibitem{xu2022star}
J.~Xu, Y.~Liu, X.~Mu, R.~Schober, and H.~V. Poor, ``{STAR}-{RIS}s: A correlated
  {T}\&{R} phase-shift model and practical phase-shift configuration
  strategies,'' \emph{IEEE J. Sel. Top. Signal Process.}, vol.~16, no.~5, pp.
  1097--1111, 2022.

\bibitem{zuo2022joint}
J.~Zuo, Y.~Liu, Z.~Ding, L.~Song, and H.~V. Poor, ``Joint design for
  simultaneously transmitting and reflecting ({STAR}) {RIS} assisted {NOMA}
  systems,'' \emph{{IEEE} Trans. Wireless Commun.}, vol.~22, no.~1, pp.
  611--626, 2022.

\bibitem{9140329}
M.~Di~Renzo, A.~Zappone, M.~Debbah, M.-S. Alouini, C.~Yuen, J.~de~Rosny, and
  S.~Tretyakov, ``Smart radio environments empowered by reconfigurable
  intelligent surfaces: {H}ow it works, state of research, and the road
  ahead,'' \emph{{IEEE} J. Sel. Areas Commun.}, vol.~38, no.~11, pp.
  2450--2525, 2020.

\bibitem{9424177}
Y.~Liu, X.~Liu, X.~Mu, T.~Hou, J.~Xu, M.~Di~Renzo, and N.~Al-Dhahir,
  ``Reconfigurable intelligent surfaces: Principles and opportunities,''
  \emph{IEEE Commun. Surv. Tutor.}, vol.~23, no.~3, pp. 1546--1577, 2021.

\bibitem{liu2023simultaneously}
Y.~Liu, J.~Xu, Z.~Wang, X.~Mu, J.~Zhang, and P.~Zhang, ``Simultaneously
  transmitting and reflecting ({STAR}) {RIS} for 6{G}: Fundamentals, recent
  advances, and future directions,'' \emph{arXiv preprint arXiv:2304.14180},
  2023.

\bibitem{9570143}
X.~Mu, Y.~Liu, L.~Guo, J.~Lin, and R.~Schober, ``Simultaneously transmitting
  and reflecting ({STAR}) {RIS} aided wireless communications,'' \emph{{IEEE}
  Trans. Wireless Commun.}, vol.~21, no.~5, pp. 3083--3098, 2022.

\bibitem{liu2022simultaneously}
Y.~Liu, X.~Mu, R.~Schober, and H.~V. Poor, ``Simultaneously transmitting and
  reflecting ({STAR})-{RIS}s: A coupled phase-shift model,'' in \emph{Proc.
  {IEEE} Intl. Conf. Commun. (ICC)}, May 2022, pp. 2840--2845.

\bibitem{wang2022simultaneously}
Z.~Wang, X.~Mu, J.~Xu, and Y.~Liu, ``Simultaneously transmitting and reflecting
  surface ({STARS}) for {T}erahertz communications,'' \emph{arXiv preprint
  arXiv:2212.00497}, 2022.

\bibitem{10052764}
T.~Wang, F.~Fang, and Z.~Ding, ``Joint phase shift and beamforming design in a
  multi-user {MISO} {STAR}-{RIS} assisted downlink {NOMA} network,''
  \emph{{IEEE} Trans. Veh. Technol.}, pp. 1--12, 2023.

\bibitem{wu2021coverage}
C.~Wu, Y.~Liu, X.~Mu, X.~Gu, and O.~A. Dobre, ``Coverage characterization of
  {STAR-RIS} networks: {NOMA} and {OMA},'' \emph{{IEEE} Commun. Lett.},
  vol.~25, no.~9, pp. 3036--3040, 2021.

\bibitem{7942128}
K.~T. Selvan and R.~Janaswamy, ``Fraunhofer and {F}resnel distances: Unified
  derivation for aperture antennas,'' \emph{IEEE Trans. Antennas Propag.},
  vol.~59, no.~4, pp. 12--15, 2017.

\bibitem{bjornson2021primer}
E.~Bj{\"o}rnson, {\"O}.~T. Demir, and L.~Sanguinetti, ``A primer on near-field
  beamforming for arrays and reconfigurable intelligent surfaces,'' in
  \emph{Proc. 55th Asilomar Conf. on Signals, Syst., and Comput.}, Oct. 2021,
  pp. 105--112.

\bibitem{9536436}
N.~J. Myers and R.~W. Heath, ``Infocus: A spatial coding technique to mitigate
  misfocus in near-field {L}o{S} beamforming,'' \emph{{IEEE} Trans. Commun.},
  vol.~21, no.~4, pp. 2193--2209, 2022.

\bibitem{miller2019waves}
D.~A. Miller, ``Waves, modes, communications, and optics: A tutorial,''
  \emph{Adv. Opt. Photonics}, vol.~11, no.~3, pp. 679--825, 2019.

\bibitem{zhang2022near}
X.~Zhang, H.~Zhang, and Y.~C. Eldar, ``Near-field sparse channel representation
  and estimation in {6G} wireless communications,'' \emph{arXiv preprint
  arXiv:2212.13527}, 2022.

\bibitem{zhang2022beam}
H.~Zhang, N.~Shlezinger, F.~Guidi, D.~Dardari, M.~F. Imani, and Y.~C. Eldar,
  ``Beam focusing for near-field multiuser {MIMO} communications,''
  \emph{{IEEE} Trans. Wireless Commun.}, vol.~21, no.~9, pp. 7476--7490, 2022.

\bibitem{wu2022multiple}
Z.~Wu, M.~Cui, and L.~Dai, ``Multiple access for near-field communications:
  {SDMA} or {LDMA}?'' \emph{arXiv preprint arXiv:2208.06349}, 2022.

\bibitem{zhang2023physical}
Z.~Zhang, Y.~Liu, Z.~Wang, X.~Mu, and J.~Chen, ``Physical layer security in
  near-field communications: What will be changed?'' \emph{arXiv preprint
  arXiv:2302.04189}, 2023.

\bibitem{zhang2022near1}
H.~Zhang, N.~Shlezinger, F.~Guidi, D.~Dardari, M.~F. Imani, and Y.~C. Eldar,
  ``Near-field wireless power transfer with dynamic metasurface antennas,'' in
  \emph{2022 IEEE 23rd International Workshop on Signal Processing Advances in
  Wireless Communication (SPAWC)}, Jul. 2022, pp. 1--5.

\bibitem{6995424}
S.~Koenig \emph{et~al.}, ``Wireless sub-{TH}z communication system with high
  data rate enabled by {RF} photonics and active {MMIC} technology,'' in
  \emph{IEEE Photon. J.}, 2014, pp. 414--415.

\bibitem{8901159}
H.~Elayan, O.~Amin, B.~Shihada, R.~M. Shubair, and M.-S. Alouini, ``Terahertz
  band: The last piece of {RF} spectrum puzzle for communication systems,''
  \emph{{IEEE} {O}pen j. Commun. Soc.}, vol.~1, pp. 1--32, 2020.

\bibitem{cui2021near}
M.~Cui, L.~Dai, R.~Schober, and L.~Hanzo, ``Near-field wideband beamforming for
  extremely large antenna arrays,'' \emph{arXiv preprint arXiv:2109.10054},
  2021.

\bibitem{miller2000communicating}
D.~A. Miller, ``Communicating with waves between volumes: evaluating orthogonal
  spatial channels and limits on coupling strengths,'' \emph{Appl. Opt.},
  vol.~39, no.~11, pp. 1681--1699, 2000.

\bibitem{shi2011iteratively}
Q.~Shi, M.~Razaviyayn, Z.-Q. Luo, and C.~He, ``An iteratively weighted {MMSE}
  approach to distributed sum-utility maximization for a {MIMO} interfering
  broadcast channel,'' \emph{{IEEE} Trans. Signal Process.}, vol.~59, no.~9,
  pp. 4331--4340, 2011.

\bibitem{grant2014cvx}
M.~Grant and S.~Boyd, ``{CVX}: Matlab software for disciplined convex
  programming, version 2.1,'' [Online]. Available:\url{http://cvxr.com/cvx},
  2014.

\bibitem{pan2020multicell}
C.~Pan, H.~Ren, K.~Wang, W.~Xu, M.~Elkashlan, A.~Nallanathan, and L.~Hanzo,
  ``Multicell {MIMO} communications relying on intelligent reflecting
  surfaces,'' \emph{{IEEE} Trans. Wireless Commun.}, vol.~19, no.~8, pp.
  5218--5233, 2020.

\bibitem{ben1997penalty}
A.~Ben-Tal and M.~Zibulevsky, ``Penalty/barrier multiplier methods for convex
  programming problems,'' \emph{SIAM J. Optim.}, vol.~7, no.~2, pp. 347--366,
  1997.

\bibitem{dinh2010local}
Q.~T. {Dinh} and M.~{Diehl}, ``Local convergence of sequential convex
  programming for nonconvex optimization,'' \emph{Recent Advances in
  Optimization and its Applications in Engineering}, Berlin, Germany: Springer,
  2010.

\bibitem{bomze2010interior}
I.~M. Bomze, V.~F. Demyanov, R.~Fletcher, T.~Terlaky, I.~P{\'o}lik, and
  T.~Terlaky, ``Interior point methods for nonlinear optimization,''
  \emph{Nonlinear Optimization: Lectures given at the CIME Summer School held
  in Cetraro, Italy, July 1-7, 2007}, pp. 215--276, 2010.

\bibitem{niu2021weighted}
H.~Niu, Z.~Chu, F.~Zhou, P.~Xiao, and N.~Al-Dhahir, ``Weighted sum rate
  optimization for {STAR}-{RIS}-assisted {MIMO} system,'' \emph{{IEEE} Trans.
  Veh. Technol.}, vol.~71, no.~2, pp. 2122--2127, 2021.

\bibitem{perera2022sum}
P.~P. Perera, V.~G. Warnasooriya, D.~Kudathanthirige, and H.~A. Suraweera,
  ``Sum rate maximization in {STAR}-{RIS} assisted full-duplex communication
  systems,'' in \emph{Proc. {IEEE} Intl. Conf. Commun. (ICC)}, May 2022, pp.
  3281--3286.

\bibitem{9935266}
Z.~Wang, X.~Mu, Y.~Liu, and R.~Schober, ``Coupled phase-shift {STAR}-{RIS}s: A
  general optimization framework,'' \emph{{IEEE} Commun. Lett.}, vol.~12,
  no.~2, pp. 207--211, 2023.

\end{thebibliography}
\bibliographystyle{IEEEtran}


\end{document}